\documentclass{article}
\usepackage[a4paper, margin=2.5cm]{geometry}
\usepackage{amsmath}

\usepackage{hyperref}
\usepackage{siunitx}

\usepackage[T1]{fontenc}
\usepackage[utf8]{inputenc}
\usepackage[english]{babel}
\usepackage[autostyle=true]{csquotes} % Required to generate
\DeclareUnicodeCharacter{2212}{-} % Needed for minus sign

\usepackage{natbib}

\usepackage{booktabs}
\usepackage{multirow}

\usepackage{tikz}

\DeclareRobustCommand\full {\tikz[baseline=-0.5ex]\draw[very thick]
  (0,0)--(0.5,0);}
\DeclareRobustCommand\dotted{\tikz[baseline=-0.5ex]\draw[very
  thick,dotted] (0,0)--(0.54,0);}

\definecolor{CustomBlue}{rgb}{0.12156862745098039, 0.4666666666666667,
  0.7058823529411765} \definecolor{CustomOrange}{rgb}{1.0,
  0.4980392156862745, 0.054901960784313725}
\definecolor{CustomGreen}{rgb}{0.17254901960784313,
  0.6274509803921569, 0.17254901960784313}
\definecolor{CustomRed}{rgb}{0.8392156862745098, 0.15294117647058825,
  0.1568627450980392)} \DeclareRobustCommand\LineBlue
{\tikz[baseline=-0.6ex]\draw[very thick, CustomBlue] (0,0)--(0.5,0);}
\DeclareRobustCommand\LineOrange {\tikz[baseline=-0.6ex]\draw[very
  thick, CustomOrange] (0,0)--(0.5,0);}
\DeclareRobustCommand\LineGreen {\tikz[baseline=-0.6ex]\draw[very
  thick, CustomGreen] (0,0)--(0.5,0);} \DeclareRobustCommand\LineRed
{\tikz[baseline=-0.6ex]\draw[very thick, CustomRed] (0,0)--(0.5,0);}

\definecolor{CustomK1}{rgb}{0.9312692223325372, 0.8201921796082118,
  0.7971480974663592}
\definecolor{CustomK2}{rgb}{0.8707745887378947,
  0.671185774711061, 0.6918315421885759}
\definecolor{CustomK3}{rgb}{0.7840440880599453, 0.5292660544265891,
  0.6200568926941761}
\definecolor{CustomK4}{rgb}{0.6672565752652589,
  0.40671838146419587, 0.5620016466433286}
\definecolor{CustomK5}{rgb}{0.5151069036855755, 0.29801047535056074,
  0.49050619139300705}
\definecolor{CustomK6}{rgb}{0.34417698402907926, 0.2047057228850617,
  0.3876123154962551}
\definecolor{CustomK7}{rgb}{0.1750865648952205,
  0.11840023306916837, 0.24215989137836502}
\DeclareRobustCommand\LineOne {\tikz[baseline=-0.6ex]\draw[very thick,
  CustomK1] (0,0)--(0.5,0);}
\DeclareRobustCommand\LineTwo
{\tikz[baseline=-0.6ex]\draw[very thick, CustomK2] (0,0)--(0.5,0);}
\DeclareRobustCommand\LineThree {\tikz[baseline=-0.6ex]\draw[very
  thick, CustomK3] (0,0)--(0.5,0);}
\DeclareRobustCommand\LineFour
{\tikz[baseline=-0.6ex]\draw[very thick, CustomK4] (0,0)--(0.5,0);}
\DeclareRobustCommand\LineFive {\tikz[baseline=-0.6ex]\draw[very
  thick, CustomK5] (0,0)--(0.5,0);}
\DeclareRobustCommand\LineSix
{\tikz[baseline=-0.6ex]\draw[very thick, CustomK6] (0,0)--(0.5,0);}
\DeclareRobustCommand\LineSeven {\tikz[baseline=-0.6ex]\draw[very
  thick, CustomK7] (0,0)--(0.5,0);}

\DeclareRobustCommand{\CovEffect}{\texttt{IMP}}

 \title{Burglary in London: Insights from Statistical Heterogeneous Spatial Point Processes}

 \usepackage{authblk} \author[1,3]{Jan Povala} \author[2,3]{Seppo Virtanen} \author[2,3]{Mark Girolami}
 \affil[1]{Department of Mathematics, Imperial College London}
 \affil[2]{Department of Engineering, University of Cambridge}
 \affil[3]{The Alan Turing Institute}
 
 \date{June 12 2020}

\usepackage{amsmath}
\usepackage{bm}

\def\figref#1{figure~\ref{#1}}
\def\tabref#1{table~\ref{#1}}
\def\Tabref#1{Table~\ref{#1}}
\def\Figref#1{Figure~\ref{#1}}

\def\secref#1{section~\ref{#1}}
\def\Secref#1{Section~\ref{#1}}
\def\eqref#1{equation~\ref{#1}}
\def\Eqref#1{Equation~\ref{#1}}
\def\1{\bm{1}}

\def\rvbeta{{\bm{\beta}}}

\def\rvpi{{\bm{\pi}}}

\def\rvtheta{{\bm{\theta}}}
\def\rvphi{{\bm{\phi}}}

\def\rvf{{\mathbf{f}}}

\def\rvy{{\mathbf{y}}}

\def\rvz{{\mathbf{z}}}

\def\ervbeta{{{\beta}}}
\def\ervpi{{{\pi}}}
\def\ervphi{{{\phi}}}
\def\ervsigma{{{\sigma}}}
\def\ervtheta{{{\theta}}}

\def\ervf{{\textnormal{f}}}

\def\ervy{{\textnormal{y}}}
\def\ervz{{\textnormal{z}}}

\def\vx{{\bm{x}}}

\def\mF{{\bm{F}}}

\def\mK{{\bm{K}}}

\def\mX{{\bm{X}}}

\DeclareMathAlphabet{\mathsfit}{\encodingdefault}{\sfdefault}{m}{sl}
\SetMathAlphabet{\mathsfit}{bold}{\encodingdefault}{\sfdefault}{bx}{n}

\newcommand{\Var}{\mathrm{Var}}

\DeclareMathOperator{\mtrace}{tr}
\newcommand{\GP}{\mathcal{GP}}

\newcommand{\BigOh}[1]{\mathcal{O}{\left({#1}\right)}}

\newcommand{\euler}{\mathrm{e}}

\begin{document}
\maketitle

\begin{abstract}
  To obtain operational insights regarding the crime of burglary in London, we
  consider the estimation of the effects of covariates on the intensity of
  spatial point patterns. Inspired by localised properties of criminal
  behaviour, we propose a spatial extension to mixtures of generalised linear
  models from the mixture modelling literature. The proposed Bayesian model is a
  finite mixture of Poisson generalised linear models such that each location is
  probabilistically assigned to one of the groups. Each group is characterised
  by the regression coefficients, which we subsequently use to interpret the
  localised effects of the covariates. By using a blocks structure of the study
  region, our approach allows specifying spatial dependence between nearby
  locations. We estimate the proposed model using Markov Chain Monte Carlo
  methods and provide a Python implementation.
\end{abstract}

\section{Introduction}
Use of statistical models for understanding and predicting criminal behaviour
has become increasingly relevant for police forces, and policymakers
\citep{felson_opportunity_1998, bowers_exploring_1999,
  predpol_predpol_2019}. While short-term forecasting of criminal activity has
been used to allocate policing resources better
\citep{taddy_autoregressive_2010,
  mohler_self-exciting_2011,aldor-noiman_spatio-temporal_2017,
  flaxman_scalable_2019,predpol_predpol_2019}, understanding the criminal
behaviour and target selection process through statistical models has a
potential to be used for designing policy changes and development programs
\citep{felson_opportunity_1998}. In this work, we consider the problem of
burglary crime in London. In the UK, burglary is a well-reported crime, but the
detection rate remains at the 10-15\% level \citep{smith_crimes_2013}. Rather
than being concerned with short-term forecasting, we focus on understanding the
effects of spatially varying explanatory variables on the target selection
through descriptive regression models. Inferences made using these models help
us understand the underlying mechanisms of burglary. The main contribution of
this work is the integration of statistical methods in spatial modelling with
the findings from the criminological literature.

%% Light intro to the problem we are dealing with
Instances of burglary can be represented as a \emph{spatial point pattern} -- a
finite or countably infinite set of points in the study region. Understanding
the intensity of the occurrences through spatially varying covariates is the
main objective of this work. The task of estimating the effects of the
covariates on the intensity can be classified as a multivariate regression
modelling, in which systematic effects of the explanatory variables are of
interest while taking into account other random effects such as measurement
errors and spatial correlation \citep{mccullagh_generalized_1998}. In the
context of spatial data, it has been widely recognised that multivariate
regression modelling techniques which do not account for \emph{spatial
  dependence} and \emph{spatial heterogeneity} can lead to biased results and
faulty inferences \citep{anselin_spatial_2000}. Spatial dependence refers to the
Tobler's first law of geography: ``everything is related to everything else, but
near things are more related than distant
things''\citep{tobler_computer_1970}. Spatial dependence manifests mostly in the
spatial correlation of the residuals of a model. In non-spatial settings, the
residuals are often assumed to be independent and identically distributed
\citep{mccullagh_generalized_1998}. Spatial heterogeneity is exhibited when the
object of interest, in our case, the intensity of a point pattern, shows
location-specific behaviour. For example, properties of the burglary point
pattern in a city centre are going to be different from the properties in a
residential area. Formalising these two concepts and incorporating them into
modelling methodology results in more accurate spatial models
\citep{anselin_spatial_2000}.

%%
%% -- Methods so far
%%
% Spatial dependence
Log-Gaussian Cox process \citep{moller_log_1998, moller_modern_2007} has been a
common approach for modelling intensity of spatial point patterns
\citep{diggle_spatial_2013, serra_spatio-temporal_2014, flaxman_fast_2015}. The
flexibility of the model is due to the Gaussian process part through which
complex covariance structures, including spatial dependence and heterogeneity,
can be accounted for. In practice, stationary covariance functions are used for
computational reasons \citep{diggle_spatial_2013}. As a result, log-Gaussian Cox
process models with stationary covariance functions handle spatial dependence
but do not account for spatial heterogeneity.

% Spatial heterogeneity
Mixture based approaches have been adopted as a way of enriching the collection
of probability distributions to account for spatial heterogeneity often observed
in practice \citep{green_introduction_2010, fernandez_modelling_2002}. Notably,
\citet{knorr-held_bayesian_2000, fernandez_modelling_2002, green_hidden_2002}
used mixtures for modelling the elevations of disease prevalence. While these
methods improve the model fit by accounting for spatial heterogeneity as wells
as spatial dependence, they provide little interpretation as to why the level is
elevated in certain areas. Also, these three methods have been tested only at a
modest scale. Following this line of work, \citet{hildeman_level_2018} proposed
a method in which each mixture component can take a rich representation that may
include covariates. Although this model is very rich in representation, the
empirical study in the paper was limited to the case of two mixtures, with one
of the components being held constant. Their study of a tree point pattern and
its dependence on soil type was carried out on a region discretised into a grid
with 2461 cells.

A very different approach to controlling for spatial heterogeneity has been
taken by \citet{gelfand_spatial_2003} who allow regression coefficients to vary
across the spatial region. The method treats the coefficients of the covariates
as a multivariate spatial process. The process is, however, very challenging to
fit and is often limited to 2 or 3 covariates
\citep[p.288]{banerjee_hierarchical_2015}. A simpler version of the same idea is
geographically-weighted regression \citep{brunsdon_geographically_1996}, where
the regression coefficients are weighted by a latent component whose properties
have to be specified a priori or learned through cross-validation.

% Tha gap - low expressivity, computational intractability - and our proposal.
Motivated by the computational challenges and limited interpretability of the
aforementioned approaches, we propose a mixture based method that takes into
account spatial dependence and is able to discover latent groups of locations
and characterise each group by group-specific effects of spatially varying
covariates. To estimate the model parameters from the limited data and to
quantify the uncertainty of the estimates, we follow the Bayesian framework.

More specifically, our approach builds upon the mixtures of generalised linear
models \citep{grun_finite_2008}, in which observations are modelled as a mixture
of different models. We cater for spatial dependence using an approach inspired
by \citet{fernandez_modelling_2002} and \citet{knorr-held_bayesian_2000}.  Our
model probabilistically assigns each location to a particular mixture component,
while imposing spatial dependence through prior information. The prior
information will suggest that locations that are close to each other are likely
to belong to the same component. We define a pair of locations to be close if
both of them are in the same block. We use the blocking structure predefined by
the census tracts, but our method allows defining custom ones. We further model
spatial dependence of the blocks using latent Gaussian processes, following
\citet{fernandez_modelling_2002}.  The posterior inferences for the individual
components consisting of regressions coefficients and the assignments of
locations to clusters are used to draw conclusions and provide insights about
the heterogeneity of the spatial point pattern across the study region.

In contrast to \citet{fernandez_modelling_2002} and \citet{green_hidden_2002},
this work considers including the covariates into each mixture component, rather
than having intercept-only components.  Compared to the approach of
\citet{hildeman_level_2018} who model the log-intensity of a point pattern as a
mixture of Gaussian random fields, our model is more constrained but provides
better scalability.

We show that the proposed methodology effectively models burglary crime in
London. By comparing our approach to log-Gaussian Cox process (LGCP), a standard
model for spatial point patterns \citep{diggle_spatial_2013}, we show that our
method outperforms LGCP and is more computationally tractable. Lastly, the
interpretation of inferred quantities provides useful criminological insights.

The rest of the paper is structured as follows.
\Secref{sec:ModellingMethodology} defines the model and details the inference
method, \secref{sec:ExperimentSetup} elaborates on our application and gives the
discussion of model choices specific to our application. The obtained results
are discussed in \secref{sec:Results}. \Secref{sec:Conclusions} concludes the
paper.

\section{Modelling methodology}
\label{sec:ModellingMethodology}

It is widely recognised that burglary crime is spatially concentrated
\citep{brantingham_notes_1981, clare_formal_2009, johnson_permeability_2010}.
It is also apparent that some areas in the study region will exhibit extreme
behaviour. For example, areas with no buildings such as parks will have no
burglary for structural reasons. To effectively model burglary, these phenomena
need to be accounted for using \emph{spatial effects}. The two important spatial
effects are \emph{spatial dependence} and \emph{spatial heterogeneity}
\citep{anselin_spatial_2000}.

For our modelling framework, we choose the Bayesian paradigm because it allows
us to formalise prior knowledge, and to quantify uncertainty in the unknown
quantities of our model. In our application, burglary data are given as a point
pattern over a fixed period of time. We discretise the point pattern onto a grid
of $N$ cells by counting the points in each cell. Although any form of
discretisation is allowed, throughout this paper, we work with a regular grid.

We model the count of points in a cell $n$, $\ervy_n$, conditioned on the
mixture component $k$ as a Poisson-distributed random variable, with the
logarithm of the intensity driven by a linear term, which is specific for each
mixture component, indexed by $k=1, \dots, K$. The linear term is a linear
combination of $J$ covariates for cell $n$, $\mX_n$, and the corresponding
coefficients, $\rvbeta_k$. The covariates need to be specified for the
application of interest and usually include the intercept.  To specify the prior
distribution for the regression coefficients, we use a prior that shrinks the
estimate towards zero. For each coefficient, we set
$\ervbeta_{k,j} \sim \mathcal{N}(0,\ervsigma^2_{k,j})$, where
$\ervsigma^2_{k,j} \sim \text{InvGamma}(1, 0.01)$. We put the uniform prior on
the intercepts, if present.

Each cell $n$ is probabilistically allocated to one of the $K$ components
through an allocation variable, $\ervz_n$, which is a categorical random
variable with event probabilities given by the mixture weights prior,
$\rvpi_{b[n]}$. The value of $\rvpi_{b[n]}$ is shared for all locations within
cell $n$'s block, $b[n]$.  The blocks for the study region are defined as
non-overlapping spatial areas spanning the whole study region. In many practical
applications, the block structure is already defined by administrative units or
census tracts. Block $b[n]$ is the block that contains the centroid point of
cell $n$. The block-specific event probabilities will express the belief that
the effect of the covariates is the same within the block unless evidence from
the observed data outweighs this information.

To model the mixture weights prior for block $b$,
$\rvpi_b = (\ervpi_{1,b},\dots,\ervpi_{K,b})$, we allow for different choices
provided that $\ervpi_{k,b}\geq 0$ and $\sum_{k=1}^{K}\ervpi_{k,b}=1$, i.e. it
is a valid probability measure.  One possible choice which also takes into
account the spatial dependence between the blocks is to model the mixture
weights prior for block $b$ and mixture component $k$ as
$$
\ervpi_{k,b}= \frac{\exp(\ervf_{k,b})}{\sum_{l=1}^{K}\exp(\ervf_{l,b})},
$$
where $\ervf_{k,b}$ is the evaluation of $f_{k}$ at the centroid of block $b$
and $f_{k}$ is an independent zero-mean Gaussian Process with hyper-parameters
$\rvtheta_k$. The prior for $\rvtheta_k$ is specified depending on the kernel
function used. We will use the squared exponential kernel throughout this work
\citep{rasmussen_gaussian_2006}.

\begin{figure}[ht]
  \begin{minipage}{.6\textwidth}
    \begin{align}
      \ervy_n | \ervz_{n}=k, \rvbeta_1, \dots, \rvbeta_K, \mX_n
      & \sim \text{Poisson}\left(\exp \left( \mX_{n} ^{\top}\rvbeta_{k}\right)\right) \nonumber \\
      \ervz_{n} | \rvpi
      & \sim \text{Cat} (\ervpi_{1, b[n]}, \dots, \ervpi_{K,b[n]}) \nonumber \\
      \rvpi_{k,b}|f_k
      & = \frac{\exp(\ervf_{k,b[n]})}{\sum_{l=1}^{K}\exp(\ervf_{l,b[n]})} \nonumber \\
      f_k | \rvtheta_k
      & \sim \GP(0, \kappa_{\rvtheta_k}(\cdot, \cdot)) \nonumber\\
      \rvtheta_k
      & \sim \text{kernel-dependent prior} \nonumber \\
      \ervbeta_{k,j} | \ervsigma^2_{k,j}
      & \sim  \mathcal{N}(0,\ervsigma^2_{k,j}) \nonumber \\
      \ervsigma^2_{k,j}
      & \sim \text{InvGamma}(1, 0.01). \nonumber
    \end{align}
  \end{minipage}
  \begin{minipage}{.3\textwidth}
    \begin{center}
      \includegraphics{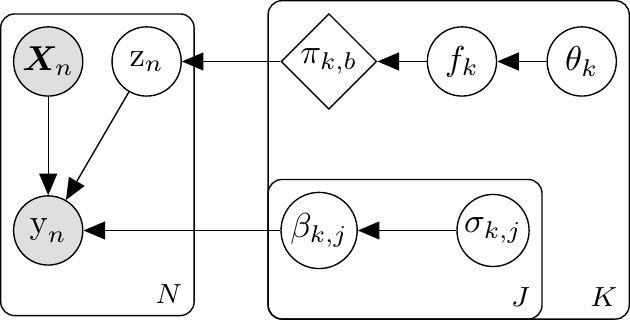}
    \end{center}
  \end{minipage}
  \caption{\label{fig:ModellingGraphicalModel}Summary of the SAM-GLM model and
    its graphical representation.}
\end{figure}

We refer to the proposed model as a \emph{spatially-aware mixture of Poisson
  generalised linear models} (SAM-GLM). The formulation is summarised in the
equation and the graphical representation shown in
\figref{fig:ModellingGraphicalModel}.
% How does our model handle heterogeneity and dependence?
In the proposed model, we handle \emph{spatial heterogeneity} using the mixture
components, each of which specifies a set of $J$ regression coefficients,
$\rvbeta_k$. \emph{Spatial dependence} is considered first within each block and
also through inter-block dependence imposed by $K$ Gaussian processes. Modelling
the spatial dependence using Gaussian processes at the block level instead of
cell level allows for more efficient estimation procedures as we discuss later.

\subsection{Excess of zeros, overdispersion}
\label{subsec:ExcessZerosOverdispersion}
Two common challenges encountered when modelling count data using standard
Poisson generalised linear models (GLM) are \emph{excess of zeros} and
\emph{overdispersion} \citep{mccullagh_generalized_1998,
  breslow_extra-poisson_1984}. The former refers to the presence of zeros that
are structural, rather than due to chance. In the context of burglary,
structural zeros occur in locations with no buildings, e.g. parks. The latter
issue refers to the situation when the variability of the observed data is
higher than what would be expected based on a particular statistical model. The
standard Poisson GLM for the burglary point pattern, a special case of our model
($K=1$), suffers from overdispersion for different specifications of the
covariates term -- see \secref{sec:OverdispersionAndExcessZeros} in the online
supplementary material. The flexibility of our proposed model can account for
the excess of zeros by identifying a low-count component to which areas of low
intensity will be assigned. Similarly, introducing mixtures can reduce
overdispersion. Two cells with similar values for the covariates, but with very
different observed counts are likely to have the same expected count under the
standard Poisson GLM. Under the mixture model, each cell would be allowed to
follow a different model.

\subsection{Inference}
Statistical inference in the Bayesian setting involves inferring the posterior
probability distribution for the quantities of interest. In this work, we choose
the Markov Chain Monte Carlo (MCMC) method to sample from the posterior
distributions \citep{gelman_bayesian_2013}.

Firstly, the scale parameter for the regression coefficients, $\sigma^2_{kj}$,
is analytically integrated out to simplify the inference (see
\eqref{eq:InferenceBetaPriorIntegrateOutSigma} in the supplementary material).
The quantities of interest are the allocation vector $\rvz$, regression
coefficient vector for each mixture component, $\rvbeta_k$, unnormalised mixture
weights priors at the centroids of the blocks, $\ervf_{k,b}$, and its
hyper-parameters.  For brevity, let $\rvbeta$ be a $K \times J$ matrix of
regression coefficients for all mixture components and each covariate, $\mX$ be
an $N \times J$ matrix of all covariates for each location, $\mF$ be a
$B\times K$ matrix such that $\mF_{b,k}=\ervf_{k, b}$, and $\rvtheta$ the vector
of kernel hyperparameters for all $f_k$'s. The unnormalised joint posterior
probability distribution is given as
\begin{equation}
  \label{eq:joint_posterior}
  p(\rvbeta, \rvz, \mF, \rvtheta | \rvy, \mX) \propto p(\rvy | \rvbeta, \mX, \rvz)p(\rvz | \mF) p(\mF | \rvtheta) p(\rvtheta) p(\rvbeta)
\end{equation}

We employ the Metropolis-within-Gibbs scheme \citep{geman_stochastic_1984,
  metropolis_equation_1953} and sample from the posterior in three steps:
\begin{enumerate}
\item We sample the regression coefficients $\ervbeta_{k,j}$ jointly for all
  $k=1,\dots,K$ and $j=1,\dots,J$. The unnormalised density of the conditional
  distribution is given as
  \begin{equation}
    p(\rvbeta | \mX, \rvy, \rvz) \propto p(\rvy | \rvbeta, \mX, \rvz) p(\rvbeta). \label{eq:MixturesInferenceBetaHmcSampling}
  \end{equation}
  \Eqref{eq:MixturesInferenceBetaHmcSampling} is sampled using the Hamiltonian
  Monte Carlo method \citep{duane_hybrid_1987}, for which efficient sampling
  schemes are available, e.g. \citet{girolami_riemann_2011}.
\item Mixture allocation is sampled cell by cell directly using the following
  equation
  \begin{equation}
    p(\ervz_n = k | \rvz^{\bar{n}}, \alpha, \mX_n, \rvbeta, \rvy, \mF) \propto p(\ervy_n|\ervz_n=k, \mX_n, \rvbeta_k)\frac{\exp(\ervf_{k,b[n]})}{\sum_{l=1}^{K}\exp(\ervf_{l,b[n]})}\label{eq:MixturesInferenceAllocSampling}
  \end{equation}
\item We sample all $K$ functions with the GP prior and their hyperparameters
  jointly using the Hamiltonian Monte Carlo. The joint posterior density is
  proportional to the expression below
  \begin{equation}
    p(\mF, \rvtheta | \rvy, \rvz) \propto \prod_{n=1}^{N}\prod_{k=1}^{K}\left(\frac{\exp(\ervf_{k,b[n]})}{\sum_{l=1}^{K}\exp(\ervf_{l,b[n]})}\right)^{I(z_n=k)} \prod_{k=1}^{K}p(f_k|\rvtheta_k)p(\rvtheta_k),
    \label{eq:MixtureInferenceGpsHmcSampling}
  \end{equation}
  where $I(\cdot)$ is the indicator function.
\end{enumerate}
For the full expansion of the conditional distributions in
equations~(\ref{eq:MixturesInferenceBetaHmcSampling}),~(\ref{eq:MixturesInferenceAllocSampling}),
and~(\ref{eq:MixtureInferenceGpsHmcSampling}), see section
\ref{sec:appendix_model_derivations} in the online supplementary material.

In terms of computational tractability,
\eqref{eq:MixturesInferenceBetaHmcSampling} takes $\BigOh{N + J}$ steps,
\eqref{eq:MixturesInferenceAllocSampling} requires $\BigOh{N \times K}$ steps,
and \eqref{eq:MixtureInferenceGpsHmcSampling} requires $\BigOh{B^3 \times K}$
steps due to matrix inversions of size $B\times B$ for each of the $K$
components.  To contrast it with a standard model for spatial point patterns,
one sample from a log-Gaussian Cox process involves matrix inversions that
require $\BigOh{N^3}$ steps \citep{diggle_spatial_2013}.  Thanks to blocking,
the inference requires inversions of smaller matrices.

\subsection{Special case: independent blocks}
\label{sec:spec-case-independent-blocks}
The model and the associated inference introduced in this section provide a very
flexible framework for modelling the spatial dependence of cells via blocks that
are also spatially dependent. However, this comes at a high cost -- inferring
posterior distribution over $K$ Gaussian processes that are combined using the
logistic function is challenging at scale as each sample requires
$\BigOh{B^3 \times K}$ operations.

If we assume that the mixture weights priors ($\rvpi_b$) for all blocks are
independent and, conditioned on $\alpha$, distributed as
\begin{equation}
  \label{eq:simulation_block_weights_prior_assumption}
  \rvpi_{b}| \alpha \sim \text{Dirichlet}(\alpha, \dots, \alpha),
\end{equation}
the inference becomes more tractable. Specifically,
\eqref{eq:MixtureInferenceGpsHmcSampling} is not needed anymore,
\eqref{eq:MixturesInferenceBetaHmcSampling} stays the same, and
\eqref{eq:MixturesInferenceAllocSampling} is replaced by
\begin{equation}
  \label{eq:simplified_model_allocation_inference}
  p(\ervz_n = k | \rvz^{\bar{n}}, \alpha, \mX_n \rvbeta, \rvy) \propto p(\ervy_n|\ervz_n=k, \mX_n \rvbeta_k)\frac{c_{b[n]k}^{\bar{n}} + \alpha}{ K \alpha + \sum_{i=1}^{K} c_{b[n]k}^{\bar{n}}}.
\end{equation}
As a result, the time complexity to take one sample from the unknown quantities
is dominated by resampling $\ervz_n$'s in
\eqref{eq:simplified_model_allocation_inference}, which can be computed in
$\BigOh{N \times K}$ steps. For the full derivation of
\eqref{eq:simplified_model_allocation_inference}, see
\secref{appendix:sssec_independent_weights_mixture_allocation_update} in the
supplementary material.

In the literature, $\alpha = 1/K$ is a recommended choice, see, e.g.,
\citet{alvares_what_2018}. This prior formulation induces sparsity and is able
to cancel out components in an overfitted mixture
\citep{rousseau_asymptotic_2011}. In the experiments we compare the trade-off
between computational complexity and modelling flexibility.

\subsection{Identifiability}
Specifying a mixture model means that the model likelihood is invariant under
the relabelling of the mixture components
\citep{celeux_computational_2000}. This issue is commonly referred to as lack of
identifiability. In the context of SAM-GLM model, $p(\rvy | \rvz, \mX, \rvbeta)$
is invariant under the relabelling of $\rvbeta_k$ and $f_k$'s, which are the
component-specific model parameters.

Exploration in high dimensional spaces is in general hard for an MCMC
sampler. As the dimension of the parameter space for the mixture model
increases, the sampler is likely to explore only one of the $K!$ possible
modes. For the sampler to switch to a different mode, it would have to get past
the area of low probability mass surrounding the chosen mode. However, note that
as the number of mixture components increases, the chance of the sampler
switching to a different mode increases as the shortest distance between a pair
of component-specific parameters is likely to decrease.

Since the identifiability issue poses a problem only for the interpretation of
the parameters, we inspect the traceplot of the Markov chain for each
identifiable parameter to assert that relabelling is not present when
interpreting the mixtures.

\section{Application: London burglary crime}
\label{sec:ExperimentSetup}

\subsection{Data description}
The methodology above has been developed to enable the analysis of our
application -- burglary in London. The data, published online by the UK police
forces \citep{police_uk_data_downloads}, are provided monthly as a spatial point
pattern over the area of $\SI{1572}{\km\squared}$ of both residential and
non-residential burglary occurrences.  The non-residential burglary refers to
instances where the target is not a dwelling, e.g., commercial or community
properties. We discretise our study area into a regular grid by counting the
number of burglary occurrences within each cell. We choose a grid for
computational reasons when comparing to competing methods (see
\secref{sec:AppendixLgcpDetails} in the supplementary material). Given our focus
on spatial modelling, we temporally aggregate the point pattern into two
datasets: the one-year dataset, starting 01/2015 and ending 12/2015, with 70,234
burglaries, and the three-year dataset, starting 01/2013 and ending 12/2015,
with 224,747 burglaries.

Our analysis uses land use data, socioeconomic census data from 2011, and points
of interest data from 2018 to estimate their effect on the intensity of the
burglary point pattern. Land use data are available as exact geometrical
shapes. The census variables are measured with respect to census tracts, called
output areas (OA). The OAs have been designed to have similar population sizes
and be as socially homogeneous as possible, based on the tenure of households
and dwelling types. Each of the 25,053 OAs in London has between 100 people or
40 households and 625 people or 250 households. The OAs are aggregated into
4,835 lower super output areas (LSOA), which in turn are aggregated into 983
middle super output areas (MSOA). An LSOA has at least 1,000 people or 400
households and at most 3,000 people or 1,200 households. For an MSOA, the
minimum is 5,000 people or 2,000 households, and the maximum is 15,000 people or
6,000 households. The points of interest data are given as a point pattern. To
project the data measured at non-grid geometries (the census and land use data)
onto the grid we use weighted interpolation. The method assumes that the data is
uniformly distributed across the OA. For cells that have an overlap with more
than one OA, we compute the value for each such cell by combining the
overlapping OAs and adjusting for the size of the overlap.

\subsection{Criminology background}
We use existing criminology studies to identify explanatory variables and
formulate hypotheses about burglary target selection. The target choice is a
decision-making process of maximising \emph{reward} with minimum \emph{effort},
and managing the \emph{risk} of being caught (a process analogous to optimal
foragers in wildlife \citep{johnson_stability_2004}). Therefore, we categorise
the explanatory variables into these three categories: reward, effort, and risk.

\subsubsection{Reward, opportunities, attractiveness}
Theoretically supported by rational choice theory \citep{clarke_modeling_1985},
offenders seek to maximise their reward by choosing areas of many opportunities
and attractive targets. Firstly, the \emph{number of dwellings} is used in the
literature as a measure of the abundance of residential targets
\citep{bernasco_how_2005, clare_formal_2009, townsley_burglar_2015,
  townsley_target_2016}. \emph{Real estate prices} and \emph{household income}
have been used in previous works as a proxy for the attractiveness of
targets. The significance of their positive effect on residential burglary
victimisation rate has been mixed and varied depending on the study region and
the statistical method used \citep{bernasco_effects_2003, bernasco_how_2005,
  clare_formal_2009, townsley_burglar_2015, townsley_target_2016}. The finding
that the effect of affluence was weak in some studies can be explained by the
fact that most burglars do not live in affluent areas and hence are not in their
awareness spaces, i.e. operating in an affluent neighbourhood is for them an
unfamiliar terrain and the risk of being caught is higher
\citep{evans_geographical_1989, rengert_use_2010}. Other measures of affluence
that have been used include \emph{house ownership rates}
\citep{bernasco_effects_2003}.

With regard to non-residential burglary, the literature is more sparse. An
analysis of non-residential burglary in Merseyside county in the UK by
\citet{bowers_exploring_1999} shows that non-residential facilities have a
higher risk of both victimisation and repeat victimisation. In particular, sport
and educational facilities have a disproportionately higher risk of being
targeted compared to other types of facilities. In the crime survey of business
owners in the UK, the retail sector is the most vulnerable to burglaries
\citep{business_crime_uk_2017_survey}. For our application, we will use points
of interest database from Ordnance Survey which include retail outlets, eating
and drinking venues, accommodation units, sport and entertainment facilities,
and health and education institutions \citep{ordnance_survey_gb_points_2018}.

\subsubsection{Effort, convenience}
Using the framework of crime pattern theory \citep{brantingham_nodes_1993} and
routine activity theory \citep{cohen_social_1979}, offenders will prefer
locations that are part of their routine activities or are convenient to them,
i.e. they are in their activity or awareness spaces. The studies performed using
the data on detected residential burglaries unanimously agree that areas
\emph{close to the offender's home} are more likely to get targeted
\citep{bernasco_how_2005, townsley_burglar_2015, menting_influence_2019,
  clare_formal_2009}. In the study based on a survey of offenders,
\citet{menting_influence_2019} argue that other awareness spaces than their
residence play a significant role in target selection. These include previous
addresses, neighbourhoods of their family and friends, as well as places where
they work and go about their recreation and leisure.

As confirmed by numerous studies, the spatial topology of the environment plays
a significant role in the choice of a target. \citet{brantingham_spatial_1975}
have shown that houses in the interior of a block are less likely to get
burgled. Similarly, \citet{townsley_burglar_2015, bernasco_how_2005} showed that
\emph{single-family dwellings} are more vulnerable to burglaries than
multi-family dwellings such as blocks of flats. \citet{beavon_influence_1994}
studied the effects of the street network and traffic flow on residential
burglary and found that crime was higher in \emph{more accessible} and
\emph{more frequented} areas. Similarly, \citet{johnson_permeability_2010} show
that main street segments are more likely to become a burglary
target. \citet{clare_formal_2009, bernasco_learning_2015} showed that the
presence of connectors such as train stations increases the likelihood of being
targeted, while the so-called barriers such as rivers or highways decrease it.

\subsubsection{Risk, likelihood of completion}
In the social disorganisation theory of crime \citep{shaw_juvenile_1942,
  sampson_community_1989}, it is argued that social cohesion induces collective
efficacy. The effect of collective efficacy on crime is twofold. First, strong
social control deters those who are thinking of committing one. Second, it
decreases the chance of a successful completion once an offender has chosen to
do so. This theory focuses on the impact that social deprivation, economic
deprivation, family disruption, ethnic heterogeneity, and residential turnover
have on the crime rates within an area. Most offenders live in disadvantaged
areas and often commit a crime in their awareness spaces (minimise effort). The
attraction to `prosperous targets' applies mostly to the local context (maximise
gain). On the other hand, when a neighbourhood has high social cohesion (also
known as `collective efficacy'), there is mutual trust among neighbours and
residents are more likely to intervene on behalf of the common good
\citep{sampson_neighborhoods_1997}.

In the context of residential burglary, \emph{ethnic diversity} has been shown
to be positively related to burglary rates \citep{sampson_community_1989,
  bernasco_how_2005, bernasco_effects_2003,
  clare_formal_2009}. \emph{Residential turnover} is another measure of
collective efficacy. Although \citet{bernasco_effects_2003} document a positive
relationship between residential turnover and the burglary rates, results in
\citet{bernasco_how_2005, townsley_burglar_2015} do not confirm that
hypothesis. \emph{Socio-economic variation} among residents has been shown to be
positively related to general crime rates
(e.g. \citet{sampson_neighborhoods_1997, johnson_testing_2015}), but it was
either not considered or shown insignificant in the studies on burglary we have
reviewed. Other indicators of social disorganisation and their effect on general
crime rates (not only burglary) are the high rate of single-parent households,
one-person households as well as younger households
\cite{bruinsma_residential_2014, sampson_neighborhoods_1997,
  andresen_place_2010}.

\subsection{Covariates selection}
\label{sec:covariates-selection}
Based on the criminological overview above and the availability of covariates,
we form four model specifications, from very rich representations to sparse
ones. \Tabref{tab:MethodModelSpecificationsCovariates} shows the covariates used
in each of the specifications.

Variables that represent density, i.e. given by the count per cell, are
log-transformed to improve the fit. For the same reason, mean household income
and mean house price are in log form. Indicators of heterogeneity are computed
using the index of variation introduced in
\citet{agresti_statistical_1978}. These include ethnic heterogeneity and
occupation variation within an area. Both are indicators of the lack of social
cohesion. Subsequently, all variables were standardised to have zero mean and
standard deviation of one.

The first specification, \emph{specification 1}, is the richest representation
and includes variables that are a proxy for the same phenomenon. For example,
both household income and house price are a measure of affluence. This choice is
deliberate as we use a shrinkage prior for the regression coefficients to choose
the most relevant variables.

The second specification, \emph{specification 2}, removes covariates that are
strongly correlated to others or lack strong evidence in the criminological
literature. We remove \emph{owner-occupied dwellings} for its strong
correlations with the house dwellings and the fraction of houses that are
detached or semi-detached. We remove \emph{house dwellings} due to high
correlation with (semi-)detached houses and stronger theoretical backing for the
latter (e.g.\citet{bernasco_how_2005}). We remove the \emph{urbanisation level}
because of little empirical evidence found in the literature. Naturally, it acts
as a proxy for where buildings are, which is accounted for to a large extent by
households and points of interest variables. We remove \emph{single-parent
  households} due to a high correlation with social housing and unemployment
rate, and the latter two being preferable indicators of social disorganisation.

In the third specification, \emph{specification 3}, we exclude the following
variables on top of those excluded in specification 2.  \emph{Median age}, as a
proxy for collective efficacy, is removed due to weak evidence in previous
studies and other measures of collective efficacy already present: ethnic and
socio-economic heterogeneity.  \emph{One-person households} and
\emph{accommodation POIs} are removed because of weak empirical evidence from
previous studies.  \emph{Mean household income} is removed due to insufficient
evidence from previous studies and an already present and more preferable
measure of affluence -- house price. \emph{Social housing} variable is removed
because of weak empirical evidence and a high correlation with unemployment.

In the last specification, \emph{specification 4}, we additionally remove
\emph{unemployment rate} due to weak empirical support from previous
studies. This specification aggregates all POIs into a single variable
(including accommodation POIs). This is to remove the strong correlations
between them. As a single variable, it signifies the level of social activity:
retail, education, entertainment, etc.

\begin{table}
  \caption{\label{tab:MethodModelSpecificationsCovariates}Models specifications
    that are used throughout the evaluation of the proposed model.}
  \centering \fbox{%
    \begin{tabular}{lllll}
      {}                                                         &            1 &            2 &            3 &            4 \\
      \midrule
      log households (count per cell)                            &  \textbullet &  \textbullet &  \textbullet &  \textbullet \\
      log retail POIs (count per cell)                           &  \textbullet &  \textbullet &  \textbullet &              \\
      log eating/drinking POIs (count per cell)                  &  \textbullet &  \textbullet &  \textbullet &              \\
      log edu/health POIs (count per cell)                       &  \textbullet &  \textbullet &  \textbullet &              \\
      log accommodation POIs (count per cell)                    &  \textbullet &  \textbullet &              &              \\
      log sport/entertainment POIs (count per cell)              &  \textbullet &  \textbullet &  \textbullet &              \\
      log POIs (all categories count per cell)                   &              &              &              &  \textbullet \\
      houses (fraction of dwellings)                             &  \textbullet &              &              &              \\
      (semi-)detached houses (fraction of dwellings)             &  \textbullet &  \textbullet &  \textbullet &  \textbullet \\
      social housing (fraction of dwellings)                     &  \textbullet &  \textbullet &              &              \\
      owner-occupied dwelling (fraction of dwellings)            &  \textbullet &    	       & 	      &   	     \\
      single-parent households (fraction of households)          &  \textbullet &    	       & 	      &   	     \\
      one-person households (fraction of households)             &  \textbullet &  \textbullet & 	      &   	     \\
      unemployment rate                                          &  \textbullet &  \textbullet &  \textbullet &              \\
      ethnic heterogeneity measure (index of variation)          &  \textbullet &  \textbullet &  \textbullet &  \textbullet \\
      occupation variation measure (index of variation)          &  \textbullet &  \textbullet &  \textbullet &  \textbullet \\
      accessibility (estimated by Transport for London)          &  \textbullet &  \textbullet &  \textbullet &  \textbullet \\
      residential turnover (ratio of residents who moved in/out) &  \textbullet &  \textbullet &  \textbullet &  \textbullet \\
      median age                                                 &  \textbullet &  \textbullet &   	      &              \\
      log mean household income                                  &  \textbullet &  \textbullet &   	      &              \\
      log mean house price                                       &  \textbullet &  \textbullet &  \textbullet &  \textbullet \\
      urbanisation index (proportion of urban area)              &  \textbullet &              &              &
    \end{tabular}
  }
\end{table}

\section{Results}
\label{sec:Results}

After discussing the modelling choices and experimental settings, we compare
SAM-GLM model to the log-Gaussian Cox process (LGCP), based on the out-of-sample
generalisation and crime hotspot prediction. For LGCP, we use the standard
formulation with a Mat\`ern covariance function (see
\secref{sec:AppendixLgcpDetails} in the online supplementary material for full
details).  Lastly, we interpret the results obtained using the proposed method
and show the relevance for obtaining criminological insights.

\subsection{Evaluation and interpretation}
\label{sec:ModelEvaluationInterpretation}

\subsubsection{Out-of-sample performance}
\label{sec:out_of_sample_performance}

Firstly, we evaluate the performance of the proposed and competing models using
the Poisson likelihood of one-period-ahead data given the model parameters
obtained from training data.  The likelihood denotes how likely the observed
data are for given parameters.  For a given sample from the posterior
distribution of the model parameters, $\rvphi^{(s)}$, the average pointwise
\emph{held-out log-likelihood} is defined as
\begin{equation}
  \label{eq:held_out_loglik}
  \text{Held-out log likelihood}=\frac{1}{N}\sum_{n=1}^{N}\log p(\tilde{\ervy}_n | \rvphi^{(s)}),
\end{equation}
where $p(\cdot)$ is the Poisson density function, $\tilde{\ervy}_n$ is the
realised next-period value.  Log-likelihood is a relative measure used for model
comparison and can only be used to compare models within the same family of
models, in our case, Poisson-based models. A higher value indicates superior
predictive power.

Next, we use the \emph{root mean square error} (RMSE) metric. It is independent
of the model and is measured at the same scale as the target variable. Given a
sample from the posterior distribution of the model parameters, $\rvphi^{(s)}$,
we obtain a sample from the joint predictive probability distribution for the
counts at all $N$ locations, $\rvy^{(s)}$, using the sampling distribution of
the data, $p(\rvy | \rvphi^{(s)})$. Then, using the realised next-period value,
$\tilde{\rvy} = (\tilde{\ervy}_1, \dots, \tilde{\ervy}_N)$, the RMSE is defined
as
\begin{equation}
  \label{eq:rmse_definition}
  \text{RMSE} = \sqrt{\frac{1}{N} \sum_{n=1}^{N}(\ervy_n^{(s)}-\tilde{\ervy}_n)^2}.
\end{equation}
A lower value of RMSE indicates a better predictive performance.

\subsubsection{Hotspot prediction}
Given that burglary is our object of interest, we also evaluate models with
respect to their ability to effectively model areas of high intensity, so-called
$hotspots$. The predictive accuracy index (PAI) and predictive efficiency index
(PEI) are two standard approaches in criminology for assessing the ability to
predict crime hotspots.

PAI, introduced by \citet{chainey_utility_2008}, assesses the ability to capture
as many crime instances as possible with the as little area as possible. For a
given size of the area to be marked as hotspots, $a$, it is defined as
\begin{equation*}
  \text{PAI} = \frac{c_a / C}{a / A},
\end{equation*}
where $A$ is the total area of the study region, $c_a$ is the number of crimes
in the flagged hotspots with the total area $a$, and $C$ is the total number of
crimes in the study region.

However, for certain types of crime that are more serious and less frequent, it
is important that each instance of crime is captured. PEI measures how effective
the model forecasts are compared to what would a perfect model predict for a
given size of the area to be marked as hotspots, $a$ \citep{hunt_crime_2016}. It
is defined as
\begin{equation*}
  \text{PEI} = \frac{c_a}{c_a^{*}},
\end{equation*}
where $c_a$ is the number of crimes in the hotspots of size $a$ flagged by the
model, and $c^*_a$ is the maximum number of crimes that could have been captured
using an area of size $a$.

In our context of a regular grid, we use both measures to compare competing
models when up to $n$ cells are flagged as hotspots. For a given $n$, a higher
value indicates better hotspot prediction ability.

\subsubsection{Interpretation of results}
Estimating the effects of different spatial covariates helps us understand the
underlying mechanisms of the point pattern.

In the mixtures of regressions literature, the interpretation of the individual
regression coefficients is of no interest, or the focus is on reporting the
regression coefficients ($\rvbeta_k$) for each component and quantifying their
uncertainty so that their significance can be assessed
\citep[ch. 8]{fruhwirth-schnatter_handbook_2019}. To further interpret the
coefficients, one could look at each mixture component specifically and
interpret the coefficients in a classical way, conditional on the partitioning
of observations. For example, for a GLM with the exponential link function,
increasing a covariate by 1 unit multiplies the mean value of the observed
variable by the exponential of the regression coefficient for that covariate,
provided other covariates are held constant. However, this approach only allows
component-specific conclusions as it depends on the distribution of the
covariate for the associated component. For example, one mixture component may
be active in areas with very small values for a specific covariate, while some
other component is active in areas with high values. Comparing regression
coefficients for that covariate across different components would not be
appropriate.

Instead, to be able to compare the covariates across mixture components, we
derive a covariate importance measure (\CovEffect{}) that is motivated by the
coefficient of determination, $R^2$.  The objective of this measure is to assess
the magnitude and the sign (positive/negative) of the effect of a covariate for
a specific mixture component on the data fit.  We measure the magnitude of the
effect for a covariate $j$ of the mixture component $k$ as the ratio of the sum
of squared residuals for the full model and the sum of squared residuals for the
same model without covariate $j$, which is then subtracted from one. For a
component $k$ and a covariate $j$,
\begin{equation}
  \text{\CovEffect{}}_{kj} = 1 - \frac{\sum_n \mathit{I}(\ervz_n=k)(\ervy_n - \hat{\ervy}_{n\tilde{\rvbeta}})^2}
  {\sum_n \mathit{I}(\ervz_n=k)(\ervy_n - \hat{\ervy}_{n\bar{\rvbeta}^j})^2},
\end{equation}
where, $\mathit{I}(\ervz_n=k)$ is the indicator function of whether cell $n$ is
allocated to component $k$, $\hat{\ervy}_{n\tilde{\rvbeta}}$ is the predicted
count using the full vector of inferred regression coefficients, and
$\hat{\ervy}_{n\bar{\rvbeta}^j}$ is the predicted count using the regression
coefficients with the $j$th coefficient set to zero. The magnitude of
\CovEffect{} is interpreted as a measure of the relative importance of the
corresponding covariate for the model fit. A value of \CovEffect{} closer to 1
represents that removing the corresponding covariate is more detrimental to
model fit.

We determine the sign of \CovEffect{} for a given covariate and a mixture
component by inspecting the distribution of the covariate for the given
component. We need to be careful with negative values as our covariates are
centred around zero and standardised. To obtain the sign, we take the mean of
the covariate across the cells that are allocated to the given component, and if
that is positive, we take the sign of the corresponding $\ervbeta_{kj}$
estimate. Otherwise, we take the negative of the sign of the $\ervbeta_{kj}$
estimate.

\subsection{Simulation study details}
For the methodology developed in \secref{sec:ModellingMethodology}, we need to
choose the grid size, blocking structure, number of mixture components ($K$) and
model specification.

\subsubsection{Model choices}
%% choice of the grid size
To choose grid size, we take into account the precision of the burglary point
pattern. The published data have been anonymised by mapping exact locations to
predefined (snap) points \citep{policeDataPage}. We follow the recommendations
in \citet{tompson_uk_2015} who assess the accuracy of the anonymisation method
by aggregating both the original and obfuscated data to areal counts at
different resolutions and looking at the difference. They show that the
aggregation at lower super output area (LSOA) level does not suffer from the
bias introduced by the anonymisation process. Therefore, for our cell size, we
approximately match an average-size LSOA to avoid the loss of precision caused
by the anonymisation process. As a result, our grid has $N=9824$ cells, each of
which corresponds to an area of $400\times400$ metres.

%% choice of the blocking structure
For the blocking structure, we take advantage of the existing census output
areas, that are designed to group homogeneous groups of households and people
together \citep{censusGeography}. Given that our grid is approximately at the
LSOA level, we choose MSOAs as the blocking structure. We assess the sensitivity
of this choice in \secref{sec:block_size_sensitivity}.

%% choice of K
The number of components, $K$, is a crucial parameter of our model. We run our
model for varying $K$ and use the performance measures introduced above to
decide on the optimal number of components. From our experience, after a certain
number of components, interpretation becomes harder while performance does not
significantly improve.

%% choice of model specification
We choose model specification based on the four options mentioned in
\secref{sec:covariates-selection}.

\subsubsection{Dependence of blocks}
\label{sec:depend-blocks-performance-justification}

{In \secref{sec:ModellingMethodology} we have proposed two possible formulations
  for the prior on the mixture weights: the multinomial logit transformation of
  $K$ Gaussian random fields and independent Dirichlet random variables. To
  assess whether assuming block dependence has a major effect on the quality of
  the model, we compare the out-of-sample performance for both variants of the
  model. For this comparison, we set the blocking scheme to MSOA, use model
  specification 4, and estimate the model on the burglary 2015 dataset. To fit
  the model with dependent blocks, we use the squared exponential kernel
  \citep{rasmussen_gaussian_2006} where we choose the lengthscale parameter by
  optimising out-of-sample RMSE using grid
  search. \Tabref{tab:dep-vs-indep-performance} shows the mean and the standard
  deviation of the samples of held-out log-likelihood and RMSE for both variants
  of the model, and for different values of $K$. The bold typeface signifies
  which method performed better for the given $K$ and for the given metric. The
  star indicates statistical significance with p-value $< 10^{-3}$ obtained from
  a two-sample t-test of samples of each metric for each variant of the model.}

  \begin{table}
    \caption{\label{tab:dep-vs-indep-performance}Model performance comparison of
      two variants of the model -- dependent blocks using the logistic transform
      of $K$ Gaussian processes, and independent blocks with Dirichlet
      prior. Reported values are a mean and standard deviation obtained from
      MCMC samples. Blocking: MSOA, training data: burglary 2015, test data:
      2016, model specification 4.}
    \sisetup{table-number-alignment=center, table-figures-decimal=3,
      detect-weight=true,detect-inline-weight=math} \centering \fbox{%
      \begin{tabular}{ l S[separate-uncertainty, table-auto-round,
          table-number-alignment = right] S[separate-uncertainty,
          table-auto-round, table-number-alignment = right]
          S[separate-uncertainty, table-auto-round, table-number-alignment =
          right] S[separate-uncertainty, table-auto-round,
          table-number-alignment = right]}
        \multirow{2}{*}{K} & \multicolumn{2}{c}{Held-out log-likelihood}  & \multicolumn{2}{c}{RMSE} \\
        \cmidrule(lr){2-3} \cmidrule(lr){4-5}  & \multicolumn{1}{r}{Independent} &
                                                                                   \multicolumn{1}{r}{Dependent}
                                               & \multicolumn{1}{r}{Independent} &
                                                                                   \multicolumn{1}{r}{Dependent} \\
        \midrule
        2 & \num{-2.607(010)} & \bfseries \num{-2.605(010)}*  &   \bfseries \num{4.999(028)}*  & \num{5.010(028)} \\
        3 & \num{-2.598(012)} & \bfseries \num{-2.593(011)}*  &   \num{4.973(036)}  & \bfseries \num{4.950(031)}* \\
        4 & \bfseries \num{-2.588(011)}* & \num{-2.606(012)}  &   \bfseries \num{4.964(034)}*  & \num{4.988(031)}
      \end{tabular}}
  \end{table}

  {The results in \tabref{tab:dep-vs-indep-performance} show that the model with
    dependent blocks does not consistently lead to improved performance. This
    indicates that block dependence structure in the burglary point pattern data
    that we consider is not a major effect. These findings highlight some
    aspects of the data structure in terms of capturing these effects and
    suggest that the point pattern data at a higher precision would be needed to
    uncover these effects, if they are present. For this reason, in the rest of
    the paper we only consider independent blocks with Dirichlet prior weights
    as described in \secref{sec:spec-case-independent-blocks}.}

\subsubsection{Identifiability}
As mentioned in \secref{sec:ModellingMethodology}, the traceplot of the
log-likelihood can be inspected for label-switching.  From our experience, the
sampler would choose one of the $K!$ modes, that are a consequence of the
likelihood invariance, and is unlikely to switch to another mode due to the high
dimensionality of the parameter space.

\subsection{SAM-GLM performance}
\begin{figure}
  \centering
  \includegraphics[width=\textwidth]{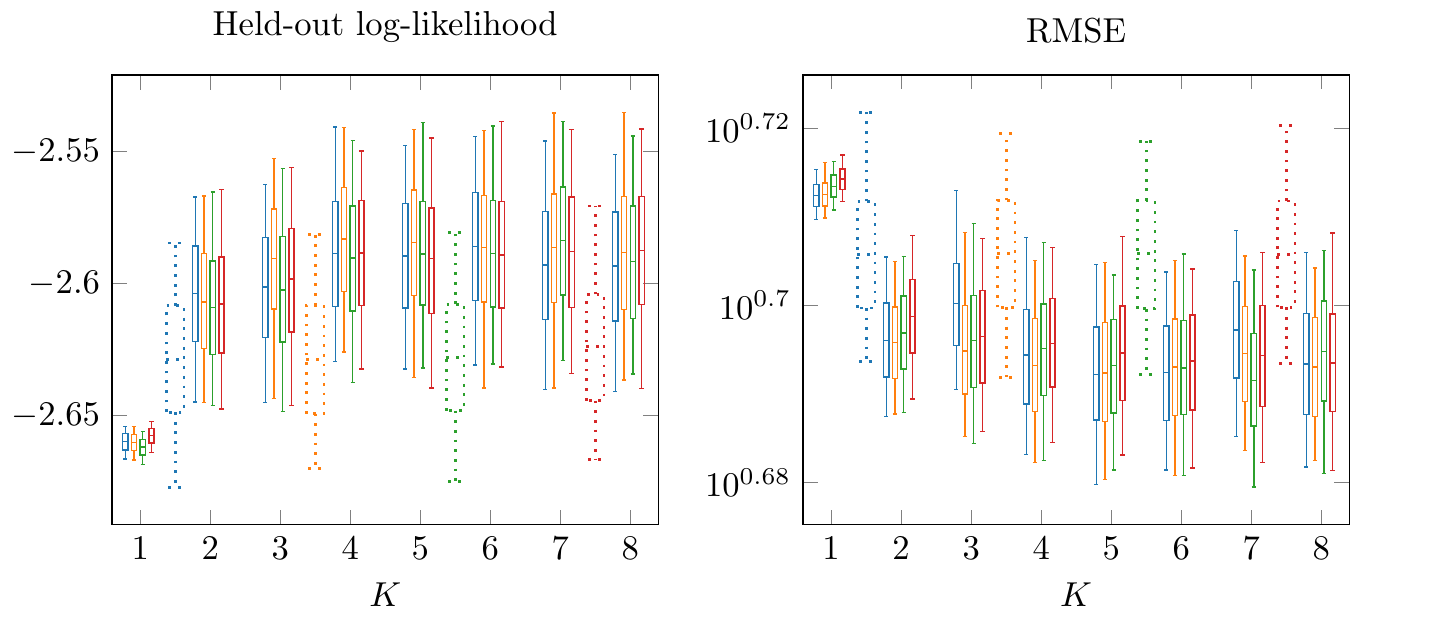}
  \caption{Evaluation of the performance of SAM-GLM (\full), compared to LGCP
    (\dotted) for the one-year dataset. Log-likelihood and root mean square
    error for the held-out data are shown for different model specifications:
    specification 1 (\LineBlue), specification 2 (\LineOrange), specification 3
    (\LineGreen), specification 4 (\LineRed). Blocking: MSOA, training data:
    burglary 2015, test data: burglary 2016. Note that the axis with the value
    of $K$ does not apply to the LGCP results.}
  \label{fig:MixturesFitPerformancePlotBurglary2015}
\end{figure}

\begin{figure}
  \centering
  \includegraphics[width=\textwidth]{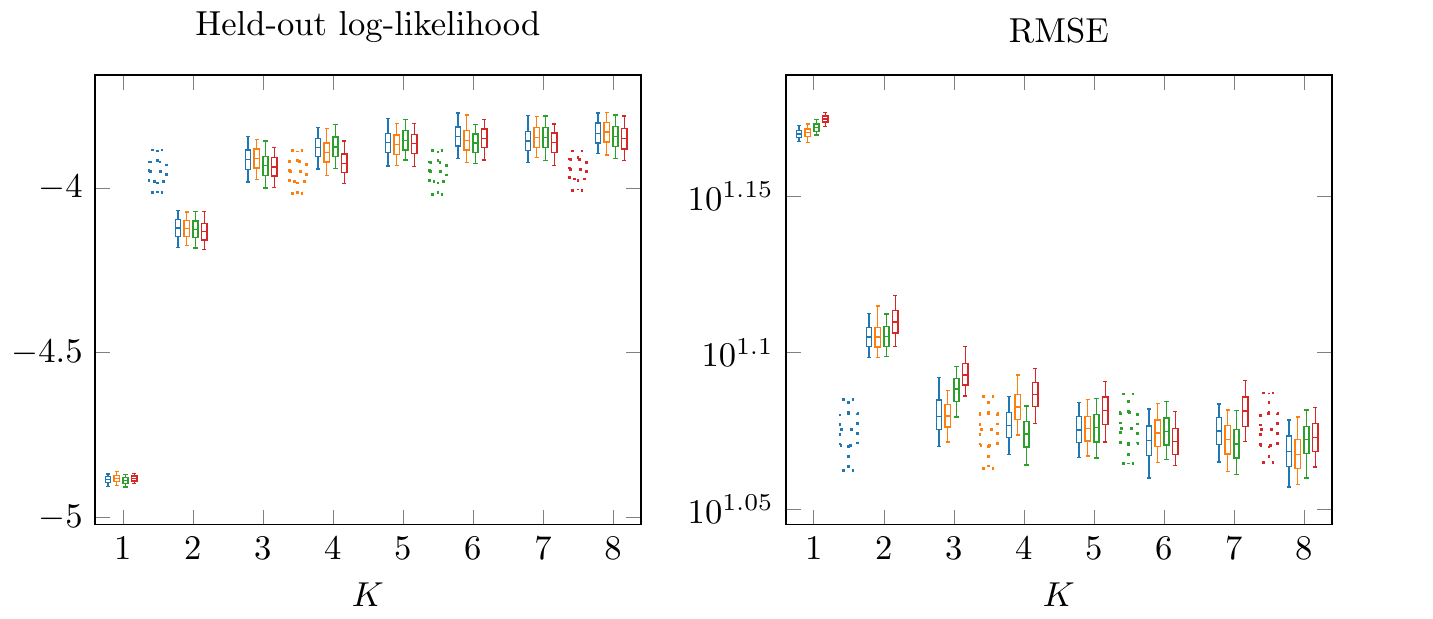}
  \caption{Evaluation of the performance of SAM-GLM (\full), compared to LGCP
    (\dotted) for the three-year dataset. Log-likelihood and root mean square
    error for the held-out data are shown for different model specifications:
    specification 1 (\LineBlue), specification 2 (\LineOrange), specification 3
    (\LineGreen), specification 4 (\LineRed). Blocking: MSOA, training data:
    burglary 2013-2015, test data: burglary 2016-2018. Note that the axis with
    the value of $K$ does not apply to the LGCP results.}
  \label{fig:MixturesFitPerformancePlotBurglary20132015}
\end{figure}

Figures \ref{fig:MixturesFitPerformancePlotBurglary2015} and
\ref{fig:MixturesFitPerformancePlotBurglary20132015} report performance for the
2015 and 2013-2015 datasets, respectively. On the left panels of the figures, we
report the box-plot of the posterior distribution of the average held-out
log-likelihood. We show the box-plot for different model specifications for both
SAM-GLM with an increasing number of components ($K$) and LGCP models. On the
right panels, we report analogous plots for the root mean square error metric
(RMSE).

For the one-year dataset, SAM-GLM model matches the predictive performance of
the LGCP model for $K=2$ components on both metrics. For the three-year dataset,
$K=3$ components are enough to match the LGCP model using the held-out
log-likelihood, but at least $K=4$ components are required for RMSE.  The extra
components required to match the performance of LGCP could be explained by the
fact that the three-year point pattern will naturally be smoother and thus
easier to interpolate non-parametrically using the Gaussian random field part of
LGCP. The probability distribution for both metrics and for all models are more
concentrated for the three-year dataset. For the one-year dataset, it is clear
that $K=2$ or $K=3$ is the optimal number of components. For the three-year
counterpart, the range between 3 and 5 components would be an appropriate
choice. For both datasets, the performance does not vary significantly for
different model specifications. Consequently, in the following sections, we
limit our attention to specification 4 due to its parsimony.

\begin{figure}
  % \centering
  \includegraphics[width=\textwidth]{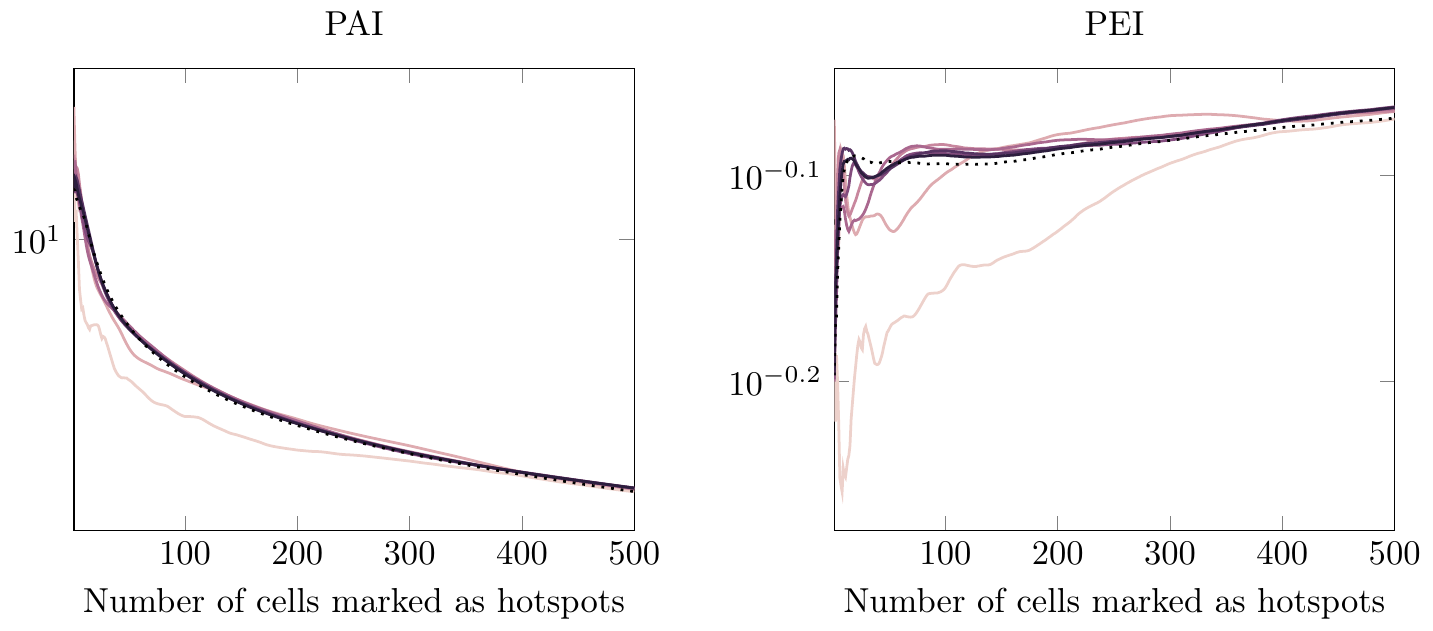}
  \caption{\label{fig:ResultsLgcpVsMixturesPaiPei2015Spec4}PAI/PEI performance
    SAM-GLM (\full) and LGCP (\dotted) models, using specification 4. For the
    SAM-GLM results, the colour of the line represents the number of components:
    $K=1$(\LineOne), $K=2$(\LineTwo), $K=3$(\LineThree), $K=4$(\LineFour),
    $K=5$(\LineFive), $K=6$(\LineSix), $K=7$ (\LineSeven). Blocking: MSOA,
    training data: burglary 2015, test data: burglary 2016, model specification:
    4.}
\end{figure}

\begin{figure}
  \centering
  \includegraphics[width=\textwidth]{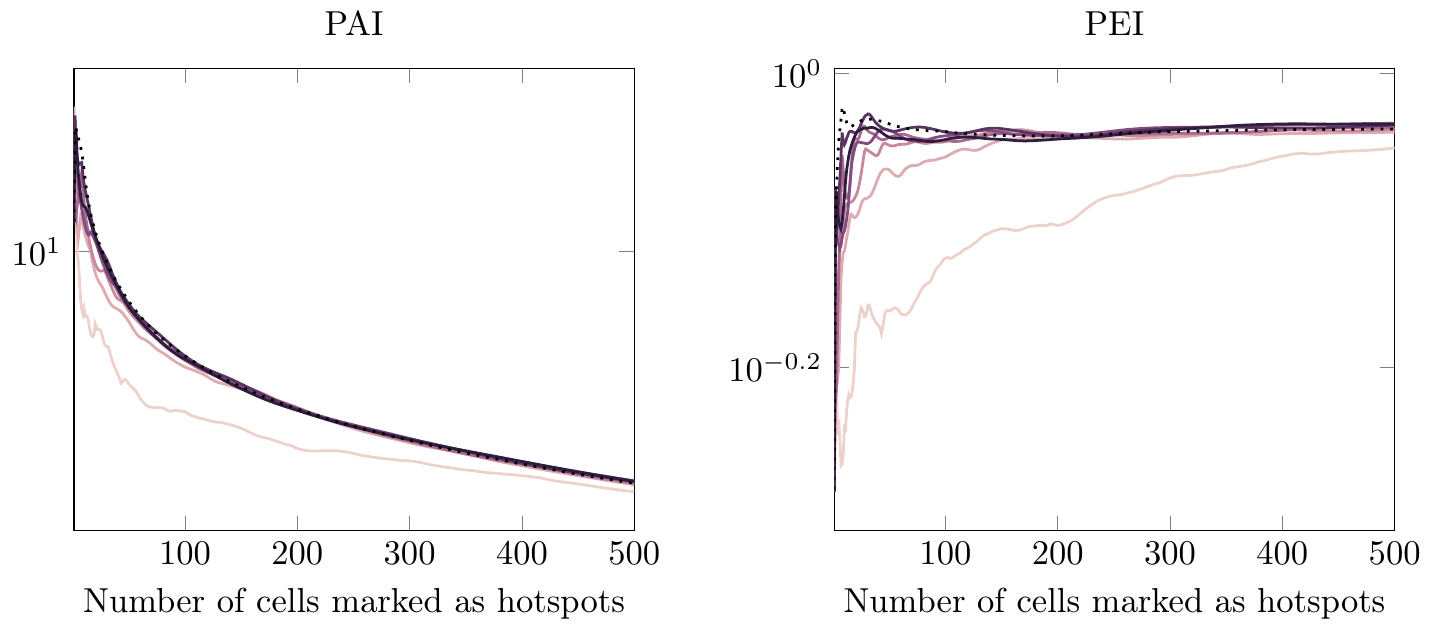}
  \caption{\label{fig:ResultsLgcpVsMixturesPaiPei20132015Spec4}PAI/PEI
    performance SAM-GLM (\full) and LGCP (\dotted) models, using specification
    4. For the SAM-GLM results, the colour of the line represents the number of
    components: $K=1$(\LineOne), $K=2$(\LineTwo), $K=3$(\LineThree),
    $K=4$(\LineFour), $K=5$(\LineFive), $K=6$(\LineSix), $K=7$
    (\LineSeven). Blocking: MSOA, training data: burglary 2013-2015, test data:
    burglary 2016-2018, model specification: 4.}
\end{figure}

While out-of-sample performance, measured by the held-out log-likelihood or
RMSE, takes into account all locations, practitioners might only be interested
in predicting crime hotspots. To this end, we evaluate PAI and PEI (see
\secref{sec:ModelEvaluationInterpretation}) as measures of hotspot
prediction. Figures \ref{fig:ResultsLgcpVsMixturesPaiPei2015Spec4} and
\ref{fig:ResultsLgcpVsMixturesPaiPei20132015Spec4} show the plots of PAI and PEI
measures for both models with specification 4, using the 2015 and 2013-2015
datasets, respectively. The plots show the score for when up to 500 cells
(around 5\% of the study region) are flagged as hotspots. Hotspots are chosen as
the $n$ cells with the highest expected value of burglaries. For the one-year
dataset, the SAM-GLM model with $K=2$ components is enough to outperform LGCP on
PEI measure when between 50 and 500 cells are flagged as hotspots.  For PAI
measure, no significant difference can be seen for $K > 2$. The results based on
the three-year data favour LGCP model when up to 150 cells are flagged as
hotspots and $K < 5$. After adding more components, the SAM-GLM performance
matches that of LGCP. When between 150 and 500 cells are flagged, $K \geq 3$
components is enough to outperform LGCP. These results are consistent with the
previous finding that outperforming LGCP on the three-year dataset requires more
components.

\subsection{Block size sensitivity}
\label{sec:block_size_sensitivity}
The proposed model requires a specification of the blocking structure for the
mixture weights prior. To assess sensitivity of this choice, we compare to
local authority districts (LAD), as well as a single block for the whole study
region. In the latter case, the model reduces to a non-spatial mixture of
Poisson GLMs. There are 946 MSOAs, and 33 LADs in the study region. The
structure is hierarchical -- multiple non-overlapping contiguous MSOAs
constitute single LAD region.

\begin{figure}
  \centering
  \includegraphics[width=\textwidth]{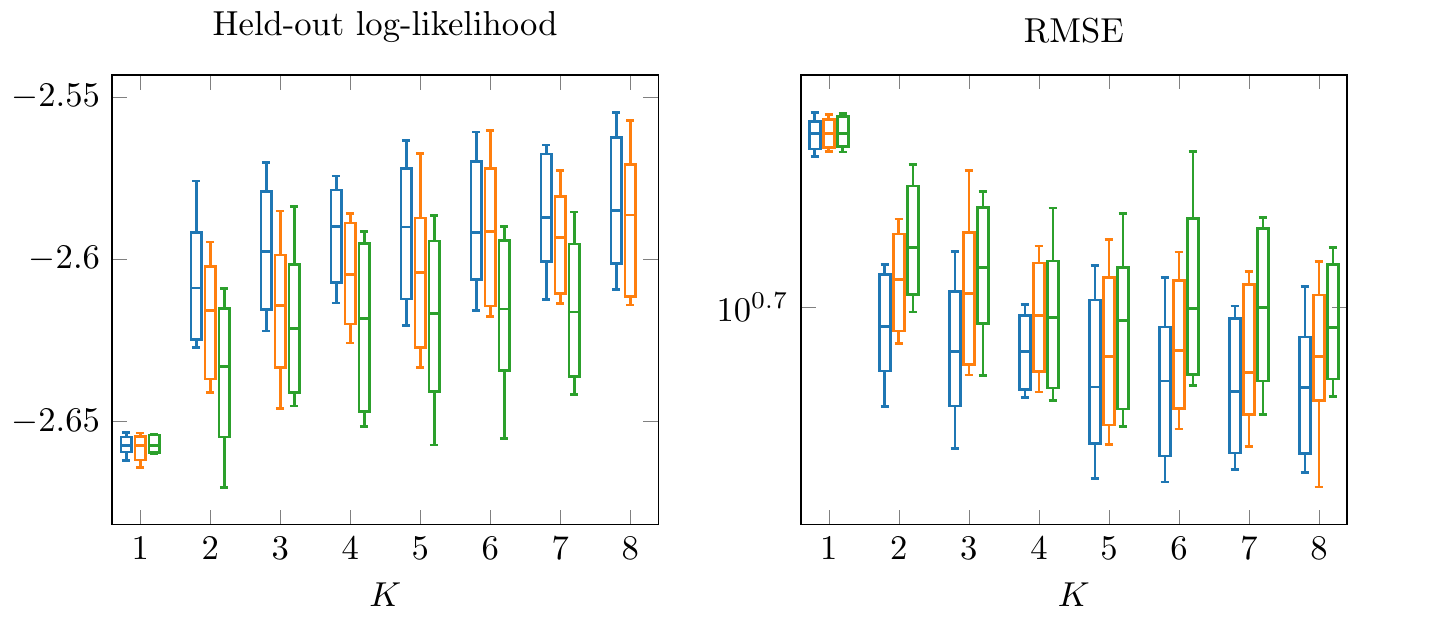}
  \caption{\label{fig:blocksize_sensitivity_analysis_12015_122015}Log-likelihood
    and root mean square error for the held-out data for different block sizes:
    MSOA(\LineBlue), LAD(\LineOrange), single block(\LineGreen). The error bars
    represent the standard deviation obtained from the respective MCMC
    samples. Training data: 2015, test data: 2016, model specification 4}
\end{figure}

\begin{figure}
  \centering
  \includegraphics[width=\textwidth]{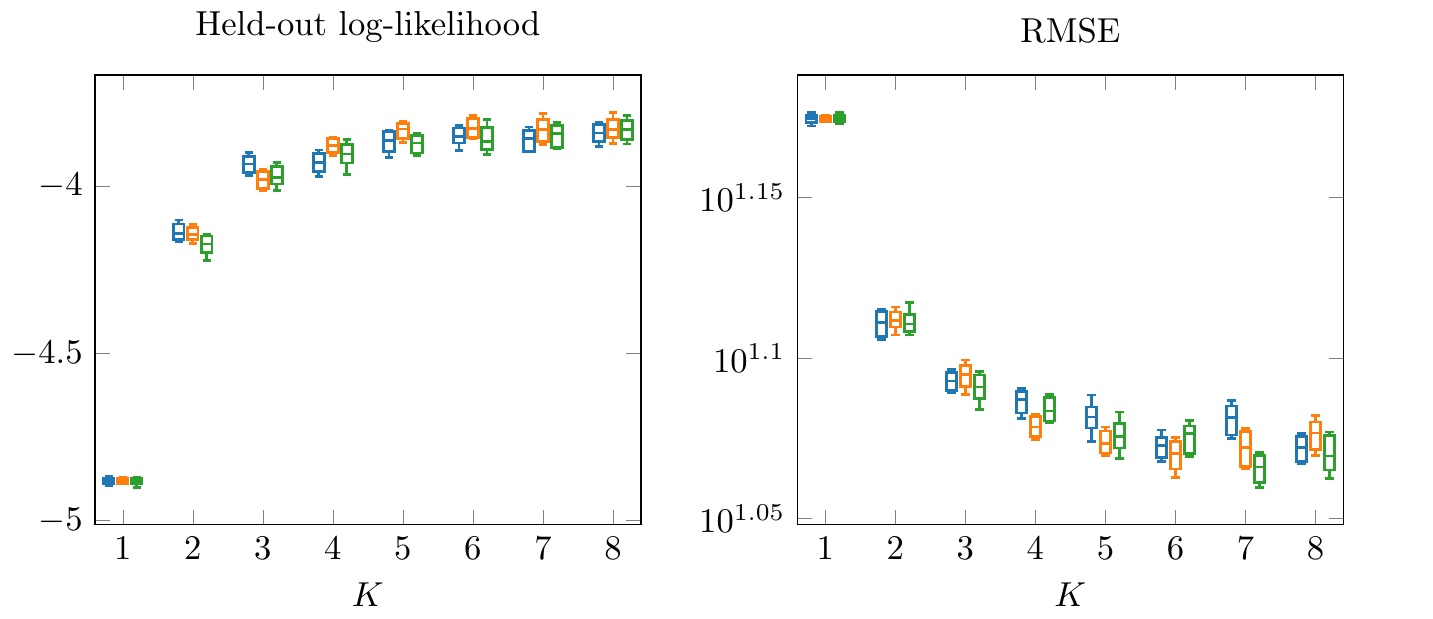}
  \caption{\label{fig:blocksize_sensitivity_analysis_12013_122015}Log-likelihood
    and root mean square error for the held-out data for different block sizes:
    MSOA(\LineBlue), LAD(\LineOrange), single block(\LineGreen).  The error bars
    represent the standard deviation obtained from the respective MCMC
    samples. Training data: 2013-2015, test data: 2016-2018, model specification
    4}
\end{figure}

Figures \ref{fig:blocksize_sensitivity_analysis_12015_122015}, and
\ref{fig:blocksize_sensitivity_analysis_12013_122015} show the box-plots of the
held-out log-likelihood and RMSE for the one-year and the three-year datasets,
respectively. The results for both metrics indicate that imposing spatial
information using more localised prior results in better out-of-sample
performance for the one-year dataset. To confirm that the difference is
statistically significant, we performed an unpaired two-sample t-test comparing
RMSE samples obtained using MSOA blocking structure to those obtained using the
LAD and single blocks, respectively. Table \ref{tab:sensitivity_analysis}
summarises the t-statistics and p-values. For the three-year dataset, there is
no evident difference, and spatial prior does not improve predictive performance
of the model. This is not surprising as the 3-year observation window will
provide more information and thus the model is less likely to overfed even if we
do not impose spatial dependence within the blocks.

\begin{table}
  \caption[Sensitivity analysis.]{\label{tab:sensitivity_analysis}Sensitivity
    analysis of block sizes. p-values comparing whether the difference in RMSE
    performance is significant. Training data: burglary 2015, test data:
    burglary 2016, specification 4.}
  \centering \fbox{%
    \begin{tabular}{lrrrr}
      \multirow{2}{*}{K} & \multicolumn{2}{c}{MSOA vs LAD}                               & \multicolumn{2}{c}{MSOA vs SINGLE}                      \\
      \cmidrule(lr){2-3} \cmidrule(lr){4-5} 
                         & \multicolumn{1}{r}{t-statistic} & \multicolumn{1}{r}{p-value} & \multicolumn{1}{r}{t-statistic} & \multicolumn{1}{r}{p-value} \\
      \midrule
      2 & -68.732 & $<\num{e-3}$ & -115.042 &$<\num{e-3}$ \\
      3 & -76.260 & $<\num{e-3}$ & -87.534  &$<\num{e-3}$ \\
      4 & -39.016 & $<\num{e-3}$ & -35.207  &$<\num{e-3}$ \\
      5 & -26.858 & $<\num{e-3}$ & -52.991  &$<\num{e-3}$ \\
      6 & -41.913 & $<\num{e-3}$ & -76.152  &$<\num{e-3}$ \\
      7 & -12.173 & $<\num{e-3}$ & -56.847  &$<\num{e-3}$ \\
      8 & -31.547 & $<\num{e-3}$ & -66.688  &$<\num{e-3}$
    \end{tabular}}
\end{table}

\subsection{Interpretation}
\begin{figure}
  \centering \fbox{
    \begin{minipage}[l]{0.60\textwidth}
      \includegraphics[width=\textwidth]{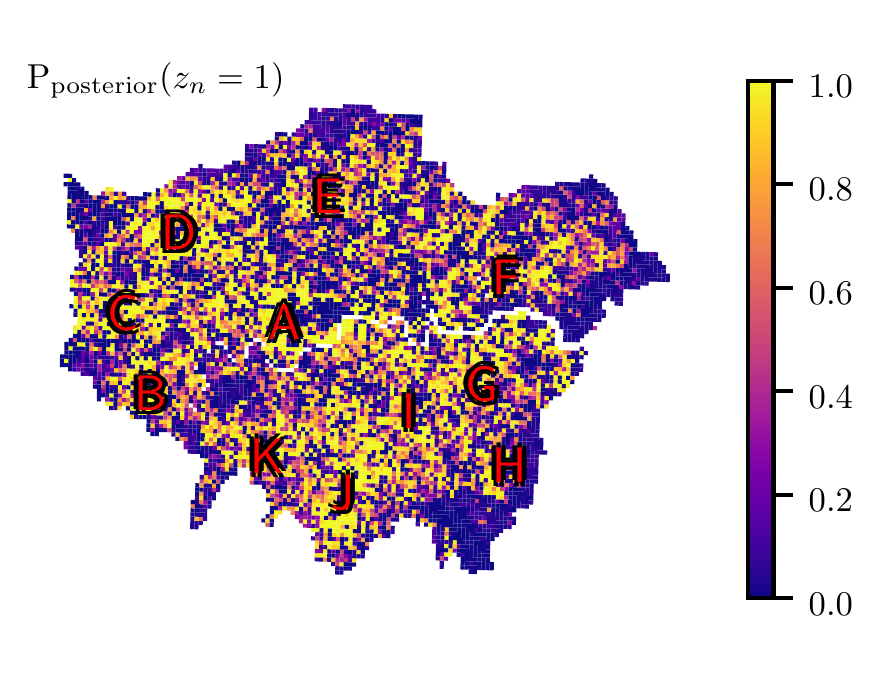}
    \end{minipage}
    \begin{minipage}[r]{0.38\textwidth}
      \resizebox{\columnwidth}{!}{%
        \begin{tabular}{lrc}
          \multicolumn{3}{c}{\CovEffect{}, component 1} \\
          \hline
          intercept                                      &       0.947 (0.001) &    + \\
          log households                                 &       0.887 (0.002) &    + \\
          log POIs (all)                                 &       0.399 (0.022) &    + \\
          accessibility                                  &       0.225 (0.024) &    + \\
          log house price                                &       0.100 (0.017) &    + \\
          ethnic heterogeneity                           &       0.083 (0.016) &    + \\
          occupation variation                           &       0.025 (0.011) &    + \\
          population turnover                            &       0.011 (0.004) &    + \\
          (Semi-)detached houses                         &       0.002 (0.002) &    + \\
          % \bottomrule
        \end{tabular}}
    \end{minipage}}
  \\
  \fbox{
    \begin{minipage}[l]{0.60\textwidth}
       \includegraphics[width=\textwidth]{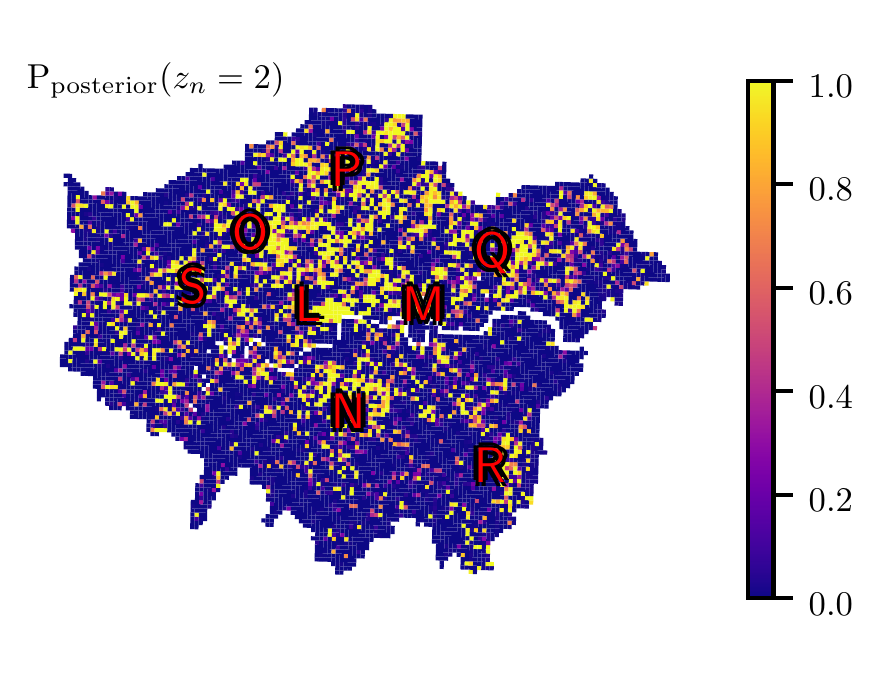}
    \end{minipage}
    \begin{minipage}[r]{0.38\textwidth}
      \resizebox{\columnwidth}{!}{%
        \begin{tabular}{lrc}
          \multicolumn{3}{c}{\CovEffect{}, component 2} \\
          \hline
          intercept                                      &       0.955 (0.001) &    + \\
          log households                                 &       0.840 (0.003) &    + \\
          accessibility                                  &       0.554 (0.012) &    + \\
          log POIs (all)                                 &       0.510 (0.017) &    + \\
          ethnic heterogeneity                           &       0.192 (0.017) &    + \\
          occupation variation                           &       0.098 (0.021) &    + \\
          population turnover                            &       0.032 (0.006) &    - \\
          log house price                                &       0.020 (0.011) &    - \\
          (Semi-)detached houses                         &       0.003 (0.002) &    + \\
          % \bottomrule &
        \end{tabular}}
    \end{minipage}}
  \\
  \fbox{
    \begin{minipage}[l]{0.60\textwidth}
      \includegraphics[width=\textwidth]{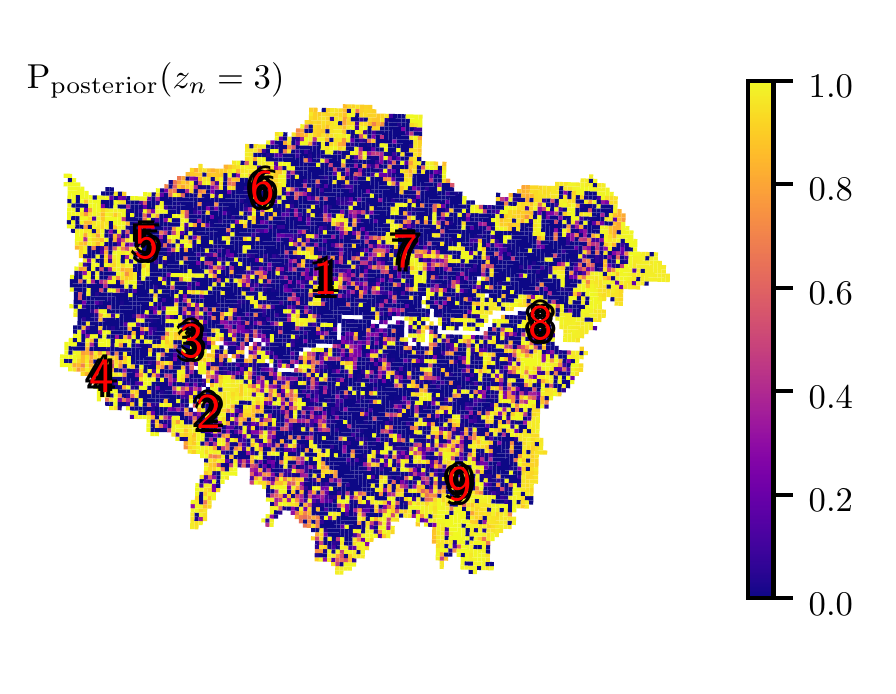}
    \end{minipage}
    \begin{minipage}[r]{0.38\textwidth}
      \resizebox{\columnwidth}{!}{%
        \begin{tabular}{lrc}
          \multicolumn{3}{c}{\CovEffect{}, component 3} \\
          \hline
          log households                                 &       0.946 (0.003) &    - \\
          intercept                                      &       0.906 (0.004) &    + \\
          log POIs (all)                                 &       0.808 (0.015) &    - \\
          occupation variation                           &       0.719 (0.060) &    + \\
          log house price                                &       0.680 (0.027) &    + \\
          accessibility                                  &       0.435 (0.086) &    + \\
          ethnic heterogeneity                           &       0.050 (0.024) &    + \\
          (Semi-)detached houses                         &       0.002 (0.007) &    + \\
          population turnover                            &       0.001 (0.007) &    + \\
          % \bottomrule &
        \end{tabular}}
    \end{minipage}}
  \caption{\label{fig:ResulsMixturesInterpretationAllocCovEffect20132015Spec4}Mixture
    model, allocations and \CovEffect{} table for each mixture
    component. Training data: 2013-2015, specification 4.}
\end{figure}

For this analysis, we choose the three-year dataset because more data will lead
to more robust inferences of the parameters. We choose specification 4 with
$K=3$ components because of its parsimony and the excellent performance shown
above -- for the three-year dataset and specification 4, there does not seem to
be a significant improvement after $K>3$ components.
\Figref{fig:ResulsMixturesInterpretationAllocCovEffect20132015Spec4} shows the
component allocation maps and the \CovEffect{} measure with the effect sign
($+/-$) for each covariate for all the three components. The allocation map for
each component shows the proportion of the MCMC samples a cell is allocated to
that component. The alphanumeric labels on the allocation plots are used in the
discussion below when referring to specific locations. \CovEffect{} is computed
for each sample and component separately and then averaged over the MCMC
samples. We also report the standard deviation of the \CovEffect{} estimate in
brackets.

% Residential component
The first component is active throughout the study region, with large clusters
around residential areas. These include areas around Kensington, Fulham, and
Shepherd's Bush (A); Hounslow, Kingston, Richmond, and Twickenham (2); Hayes and
Southall (C); Harrow and Edgware (D); East Barnet, Enfield, Walthamstow, Wood
Green (E); Barking and Dagenham (F); Bexley (G); Orpington (H); Bromley (I);
Croydon, and Purley (J); New Malden, and Morden (K). In this component, the
number of households and points of interest have the strongest effect (excluding
the intercept) -- burglaries happen where targets are. Accessibility has also
been inferred as an important covariate, consistent with the past criminological
studies. In this component, house price is inferred as having a positive effect
on the intensity of burglary, suggesting that offenders choose attractive
targets. The positive effect of ethnic heterogeneity confirms the hypothesis
from the social disorganisation theory. The other indicators of social
disorganisation -- occupation variation, residential turnover -- are weaker but
are consistent with the existing criminology literature. House price as a
measure of reward and the proportion of houses that are detached and
semi-detached have low \CovEffect{} value.

% High-intensity component
Component 2 is active in the city centre and in the high streets of
neighbourhoods: Soho, Mayfair, Covent Garden, Marylebone, Fitzrovia (L);
Shoreditch and Stratford (M); Streatham and Tooting Bec (N); Wembley, and Brent
(O); Enfield, Hampstead (P); Romford (Q); Orpington (R); Wembley, Harrow
(S). Burglary rates in these locations are largely driven by points of interest
and households. Compared to the first component (residential), the magnitudes of
\CovEffect{} values for these covariates are different - points of interest are
more important for this component, and the number of households is more
important for the first component.  Accessibility measure is inferred to have
high importance in this component. This measure is high in the city centre and
around the high streets, which are usually well-connected to the public
transport system. This confirms findings from crime pattern theory and routine
activity theory which suggest that offenders choose locations that are part of
their usual routine and in their awareness spaces. Ethnic heterogeneity and
occupation variation have strong positive effect and signify the lack of social
cohesion. Unexpectedly, our model infers a negative relationship between
residential turnover and burglary intensity. Association of high residential
turnover with the reduced risk of burglary apprehension has been shown as
significant in only a few studies and was limited to \emph{residential} burglary
\citep{bernasco_effects_2003, bernasco_how_2005, townsley_burglar_2015}. Areas
that are less residential such as high streets have a higher proportion of
flats. Dwellings with shared premises such as flats have been shown to less
likely become a target than one-household buildings
\citep{beavon_influence_1994}. Another possible reason could be the staleness of
the data for the covariates which are taken from the 2011 census. Also, house
price has been inferred to have a negative effect, i.e. more affluent locations
are less likely to get targeted. This is contrary to the first component.  A
possible explanation mentioned in previous studies is that offenders often live
in disadvantaged areas and choose targets within their awareness spaces, which
are less likely to be affluent areas \citep{evans_geographical_1989,
  rengert_use_2010}.

% Low-intensity component
The last component is active in the areas of low intensity. These include Hyde
Park, Regent's Park, Hampstead Heath (1); Richmond and Bushy parks (2); Osterley
Park and Kew botanic gardens (3); Heathrow airport (4); RAF Northolt, and parks
near Harrow (5); Edgware fields (6); Lee Valley (7); industrial zone in Barking
and Rainham Marshes (8); parks around Bromley and Biggin Hill airport (9); and
other non-urban areas located on the edges of the map. This component explains
locations with little criminal activity, signified by negative \CovEffect{} for
the number of households and points of interest. Occupation variation, as a
measure of socioeconomic heterogeneity, is strongly positive, which would
support the hypothesis from social disorganisation theory. However, this is more
likely due to the very low population in those areas which results in high
occupation variation measure. Accessibility measure also has a positive effect
on burglary rates in these locations. This is expected and in line with the
hypotheses from the crime pattern theory. Other covariates have very small
\CovEffect{} values.

The allocation of cells partitions the map into three clusters. By aggregating
the number of observed crimes that occurred in each cluster we get that
components 1, 2, and 3 cover 46\%, 42\%, 12\% of all burglaries during the
2013-2015 period, respectively. Official aggregated police data for this period
make the split of 64\% and 36\% for residential and non-residential burglary
\citep{police_uk_data_downloads}. Our inference agrees that there is more
residential burglary than non-residential burglary and that approximately
35-45\% of burglaries are non-residential. It is unclear whether the crime in
low-count areas, which according to our model accounts for 12\%, is residential
or non-residential.

The support for the presence of spatial heterogeneity is further given by
inspecting the inferences made by the LGCP model (for LGCP details see
\secref{sec:AppendixLgcpDetails} in the supplementary material). The left panel
of \figref{fig:LgcpInspection2015Spec4} shows standard deviations of the
marginal posterior distributions of the Gaussian random field component
($f$). It is clear that the variance of the field component is clustered, where
the regions with higher values are easily identifiable as those less
urbanised. In contrast, SAM-GLM model has pickled up this heterogeneity by
allowing a separate component for it (see component 3 in
\figref{fig:ResulsMixturesInterpretationAllocCovEffect20132015Spec4}). The right
panel of \figref{fig:LgcpInspection2015Spec4} shows \CovEffect{} computed for
all components of the LGCP model. \CovEffect{} measure for the field component
of the model is computed by treating it as a covariate with the coefficient
equal to one. The \CovEffect{} value for the latent field component is the
third-highest, after the intercept and the number of households. A large
contribution from the latent component indicates that the linear term in the
Poisson regression model cannot on its own sufficiently explain the variation in
the intensity of burglary.

\begin{figure}
  \centering \fbox{
    \begin{minipage}[l]{0.60\textwidth}
      \includegraphics[width=\textwidth]{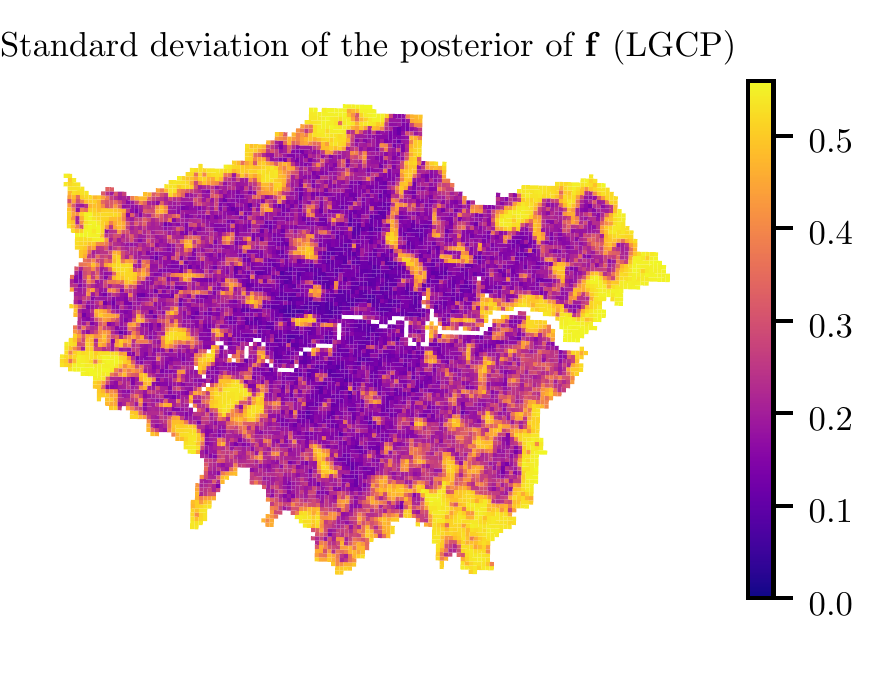}
    \end{minipage}
    \begin{minipage}[l]{0.38\textwidth}
      \resizebox{\columnwidth}{!}{%
        \begin{tabular}{lrc}
          % \toprule
          \multicolumn{3}{c}{\CovEffect{}, LGCP} \\
          \hline
          intercept              &  0.953 (0.001) &    + \\
          log households         &  0.937 (0.002) &    + \\
          field                  &  0.810 (0.005) &    + \\
          log POIs (all)         &  0.728 (0.013) &    + \\
          accessibility          &  0.103 (0.058) &    + \\
          occupation variation   &  0.097 (0.052) &    + \\
          log house price        &  0.089 (0.043) &    + \\
          ethnic heterogeneity   &  0.061 (0.029) &    + \\
          population turnover    &  0.002 (0.027) &    + \\
          (Semi-)detached houses &  0.000 (0.027) &    - \\
        \end{tabular}}
    \end{minipage}}
  \caption{\label{fig:LgcpInspection2015Spec4}Left: Standard deviation of the
    posterior distribution of the latent field, $\rvf$, of the LGCP model. It is
    clear that, it is clustered and the elevated levels correspond to non-urban
    locations, airports, and parks (see the discussion above). Right:
    \CovEffect{} measure for the component of the LGCP model. For both panels,
    training data: 2013-2015, model specification: 4.}
\end{figure}

\subsection{Overdispersion, excess of zeros}
\label{sec:results_overdispersion_excess_zeros}
The discussion of the inferences above shows that our model effectively handles
excess of zeros by allocating low-count cells (non-urban areas) its own cluster,
which has its own regression coefficients. Similarly, the proposed mixture model
is able to reduce the overdispersion problem that is present in the standard
Poisson GLM model (the special case of SAM-GLM, with $K=1$). The mixture model
may allocate each cell to a cluster that better describes the burglary count in
that location.  Inspecting the Pearson $\chi^2$ statistic
($\chi^2 = \sum_{i=1}^{N}\frac{(\text{Observed}_i -
  \text{Expected}_i)^2}{\text{Expected}_i}$) provides supporting evidence for
this. Introduction of two extra components has resulted in the 81\% decrease,
from \num{106942.43} to \num{20028.99}, showing a better model fit. This is
further confirmed by a scatter plot of expected vs observed counts for the
Poisson GLM model and the proposed model with $K=3$ as shown in
\figref{fig:handling_overdispersion} in the supplementary material.

\section{Conclusions}
\label{sec:Conclusions}
Spatial point patterns on large spatial regions, such as metropolitan areas,
often exhibit localised behaviour. Motivated by this, we propose a mixture model
that accounts for spatial heterogeneity as well as incorporates spatial
dependence. Each component of the mixture is a model in itself, and thus allows
for different locations to follow a different model, e.g. in the urban context,
less-urbanised locations can assume a different model from the city centre. Each
component is an instance of the generalised linear model (GLM) which includes
covariates. We account for spatial dependence through the mixture allocation
part. The allocation of each location to one of the components is informed by
both the data and the prior information. By utilising existing blocks structure,
or defining a custom one, the prior supports locations within the same block to
come from the same component. This formulation attempts to find the right
balance between the ability to model sharp spatial variations and borrowing
statistical strength for locations within the same block.  Additionally, the
model allows for spatial dependence between the blocks. Following the Bayesian
framework, we present a Markov Chain Monte Carlo sampler to infer the posterior
distributions. Inspection of the posterior distributions of the model parameters
allows us to learn new insights about the underlying mechanisms of the point
pattern.

Our results show that London burglary data are effectively modelled by the
proposed method. Using out-of-sample and crime hotspot prediction evaluation
measures, we show our model outperforms log-Gaussian Cox process (with Mat\`ern
covariance function) that is the default model for point processes and is more
computationally tractable.

The focus of this work on burglary crime does not limit the potential uses of
the proposed model. We believe that the model can be applied in a wider setting
of analysing spatial point patterns that may show localised behaviour and
heterogeneity.

Future analysis could consider several directions not explored in this work.
Firstly, our inference scheme for the model with block dependence produces an
$\BigOh{B^3 \times K}$ algorithm. To reduce this complexity, one could consider
$K$ level sets of a single Gaussian random field for mixture weights, instead of
$K$ Gaussian fields, thus reducing dimensionality
\citep{hildeman_level_2018,fernandez_modelling_2002}. Another approach is
assuming Markovian structure of the Gaussian random fields, resulting in sparse
computational methods\citep{rue_gaussian_2005}. A different approach is
considering inference schemes that are less computationally demanding than MCMC
such as variational methods \citep{jordan_introduction_1999}. Secondly,
different options for specifying the term that involves covariates could be
explored. One could consider forcing certain covariates to share the
coefficients across all components if there is a strong prior belief for doing
so. Another possible area of investigation is spatially varying coefficient
processes method, proposed by \citet{gelfand_spatial_2003}.

\section{Acknowledgements}
We thank the Associate Editor and the anonymous referees for their helpful
comments and suggestions. All authors were supported by the EPSRC grants
EP/P020720/1 and EP/P020720/2 for Inference, COmputation and Numerics for
Insights into Cities (ICONIC, \url{https://iconicmath.org/}). Jan Povala was
also supported by EPSRC (EP/L015129/1). Mark Girolami was supported by EPSRC
grants EP/T000414/1, EP/R018413/2, EP/R034710/1, EP/R004889/1, and a Royal
Academy of Engineering Research Chair in Data Centric Engineering. Parts of this
work were carried out while Jan Povala was visiting Strategic Insights Unit at
the Metropolitan Police Service in London. The authors are also grateful to
Louis Ellam for insightful comments and suggestions, and to Pavol Povala for his
feedback on the drafts of the manuscript.

\section{Implementation and supplementary material}
\label{sec:implementation}
The source code that implements the methodology and reproduces the experiments
is available at \url{https://github.com/jp2011/spatial-poisson-mixtures}. The
supplementary material with mathematical derivations and supporting figures is
available online.

\bibliography{references}

\appendix
\section{Poisson regression model: excess of zeros, overdispersion}
\label{sec:OverdispersionAndExcessZeros}
In this section we demonstrate that the standard Poisson regression
\citep{mccullagh_generalized_1998} is not a suitable model for the London
burglary point pattern.

\begin{figure}
  \centering
  \begin{minipage}[c]{.48\textwidth}
    \includegraphics[width=\textwidth]{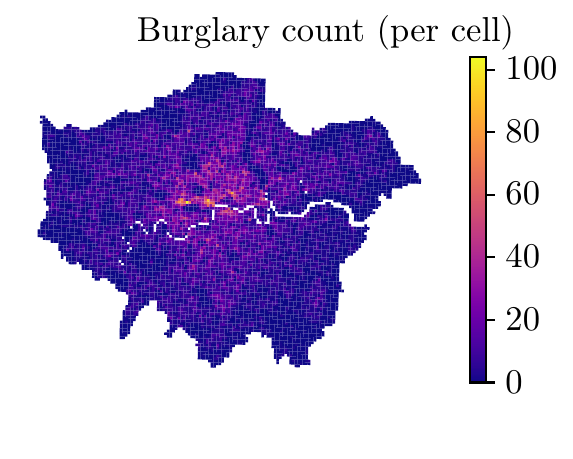}
  \end{minipage}%~
  \begin{minipage}[c]{.48\textwidth}
    \includegraphics[width=\textwidth]{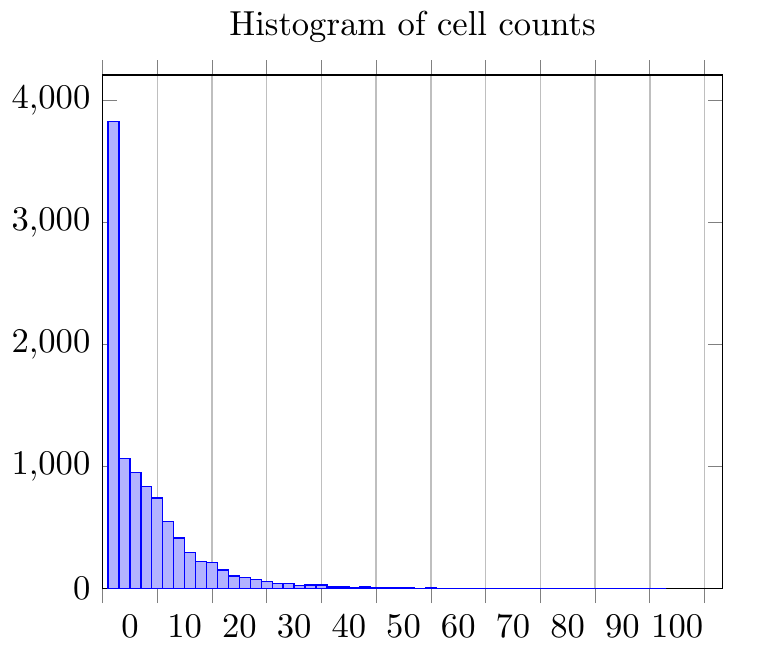}
  \end{minipage}
  \caption{\label{fig:AppendixCountsHistogram} Observed count on the map (left)
    and the corresponding histogram (right) for the point pattern of burglary
    aggregated over the grid for the time period 1/2015-12/2015.}
\end{figure}

Firstly, the dataset consists of areas with no buildings in it, e.g. parks,
airports, which results in counts equal to zero due to structure rather than due
to chance. This is further supported by the plot of the observed count and the
corresponding histogram, both shown in
\figref{fig:AppendixCountsHistogram}. This phenomenon is often referred to as
\emph{excess of zeros}.

Secondly, we fit Poisson GLM with all four specifications of covariates to the
2015 burglary dataset, as described in the paper. Then we use the overdispersion
test proposed in \citet{cameron_regression-based_1990}, and implemented in the
AER package \citep{kleiber_applied_2008}. For the standard Poisson GLM model,
$\Var(\ervy_n)=\mu_n$. The overdispersion test uses it as the null hypothesis,
where the alternative is $\Var(\ervy_n)=\mu_n + c \times g(\mu_n)$, where
$g(\cdot)$ must be specified. For our test, we choose
$g(\cdot)=1$. \Tabref{tab:OverdispersionTest} shows the estimated $c$ values and
the p-values for each estimate, given that null hypothesis is $c=0$. The data
clearly show the presence of overdispersion in all four models.

\begin{table}
  \caption{\label{tab:OverdispersionTest}Overdispersion test for Poisson GLM
    model.}
  \centering \fbox{%
    \begin{tabular}{lrr}
      Specification & $c$ & p-value\\
      \hline
      1  & 1.905 & 2.2e-16 \\
      2  & 1.897 & 2.2e-16 \\
      3  & 1.910 & 2.2e-16 \\
      4  & 1.911 & 2.2e-16
    \end{tabular}
  }
\end{table}

\subsection{Poisson regression vs SAM-GLM}
\label{sec:poiss-regr-vs-samglm}

\begin{figure}
  \centering%
  \includegraphics[width=\textwidth]{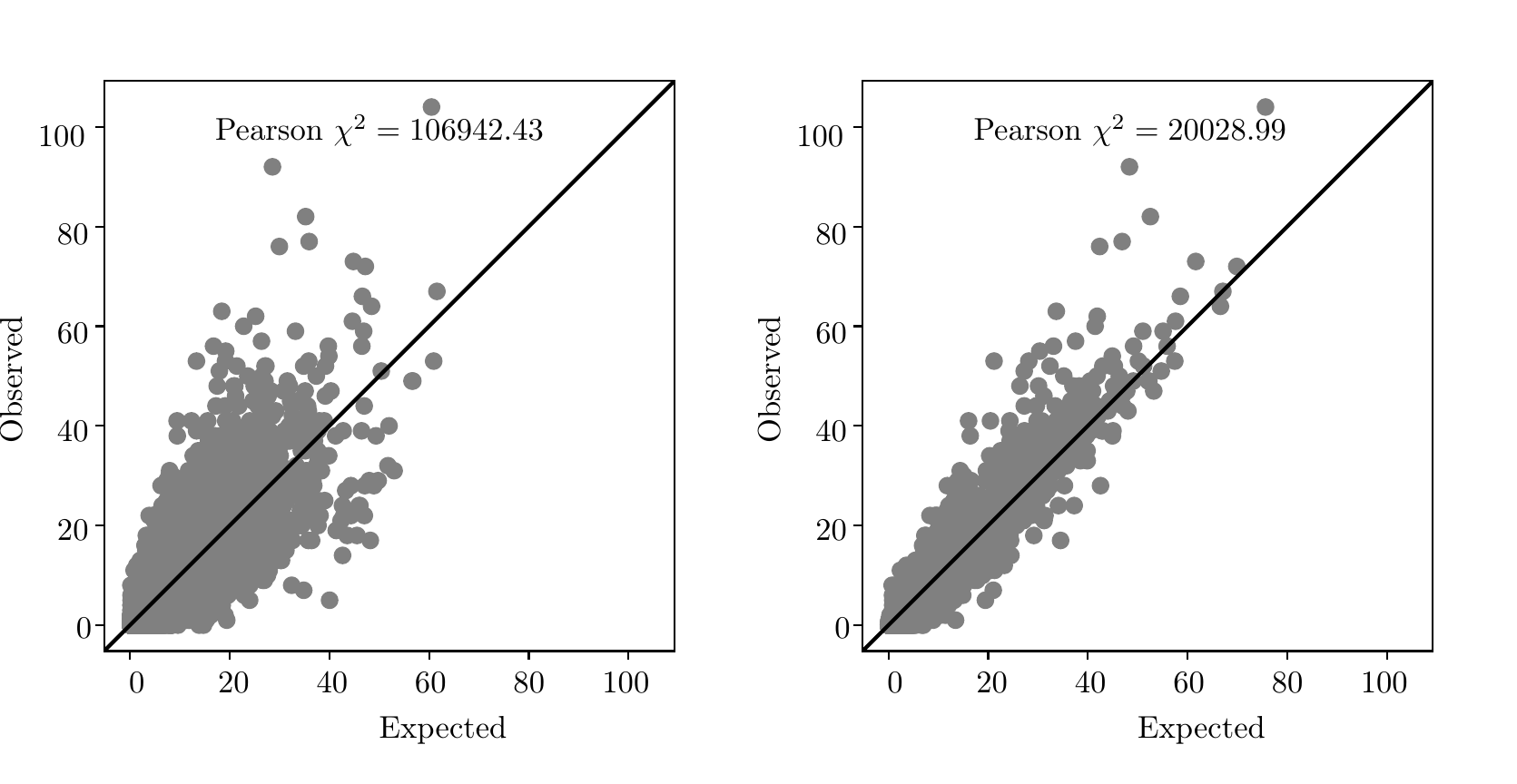}
  \caption[Overdispersion plot]{\label{fig:handling_overdispersion}Scatter plot
    of predicted counts vs observed counts (training data) for the Poisson GLM
    model (left), and SAM-GLM K=3 (right). Blocking: MSOA, training data: 2015,
    using specification 4.}
\end{figure}

\Figref{fig:handling_overdispersion} shows the scatter plot of expected vs
observed counts for the Poisson regression model (SAM-GLM with $K=1$) and the
proposed model with $K=3$. It is evident from the plot that adding extra
components to the standard Poisson regression reduces the overdispersion issue.

\section{Log-Gaussian Cox process}
\label{sec:AppendixLgcpDetails}

Dicretising the spatial domain to a regular grid, the full Bayesian formulation
of the model is given as follows:
\begin{eqnarray}
\ervy_n | \rvbeta, \rvf, \mX
& \sim & \text{Poisson} \left( \exp( \mX_{n}^\top \rvbeta  + \ervf_n) \right) \\
f(\cdot)| \rvtheta
& \sim & \GP\left(0, k_{\rvtheta}(\cdot, \cdot)\right)                         \\
\ervbeta_j
& \sim & \mathcal{N}(0,\ervsigma^2_{j})                                        \\
\ervsigma^2_{j}
& \sim & \text{InvGamma}(1, 0.01) \\
\rvtheta
& \sim & \text{weakly-informative log-normal prior},
\end{eqnarray}
where $n=1, \dots, N$ is the index over the cells on the map, $j=1, \dots, J$ is
the index over the covariates, $f()$ is a zero-mean Gaussian process with
covariance function $k_{\rvtheta}(\cdot, \cdot)$, and hyperparameters
$\rvtheta$, $\ervf_n$ is the value of $f(\cdot)$ in the centre of cell $n$,
$\mX_n$ is the vector of the covariates at cell $n$, and $\rvbeta_{j}$ is the
$j$th regression coefficient with a scale hyperparameter $\sigma^2_{kj}$. A
plain Poisson generalised linear model (GLM) formulation assumes no spatial
correlation, i.e. $\ervf_n=0$ for all $n$. Compared to the Poisson GLM model,
LGCP allows for modelling the variation in the intensity that cannot be
explained by the covariates $\mX$.

In order to allow for Kronecker product factorisation of the covariance matrix
of the Gaussian process, we specify $k_{\rvtheta}(\cdot, \cdot)$ as a product of
two Mat\'ern covariance functions, one for the easting (E) coordinate, the other
for the northing (N) coordinate. Mat\'ern covariance function is a standard
choice in spatial statistics as it allows specifying smoothness of the function
\citep{stein_interpolation_1999}. It is given as follows
\begin{align}
k _ { \mathrm { Matern } } (\vx, \vx^\prime) = \frac { 2 ^ { 1 - \nu } } { \Gamma ( \nu ) } \left( \frac { \sqrt { 2 \nu } |\vx - \vx^\prime| } { \ell } \right) ^ { \nu } K _ { \nu } \left( \frac { \sqrt { 2 \nu } |\vx - \vx^\prime| } { \ell } \right),
\end{align}
where $\ell$ is the characteristic lengthscale, $\nu$ is the smoothness
parameter, and $K _ { \nu }$ is a modified Bessel function
\citep{rasmussen_gaussian_2006}. It can be shown that that the Gaussian
processes with Mat\'ern covariance functions are $k$-times mean-square
differentiable if and only if $\nu > k$. \citet{abramowitz_handbook_1965} show
that if $\nu$ is a half-integer, i.e. for an integer $p$,
$\nu = p + \frac{1}{2}$, the covariance function becomes especially simple,
giving
\begin{align}
k _ { \ell, \nu = p + 1 / 2 } (\vx, \vx^\prime) = \exp \left( - \frac { \sqrt { 2 \nu } |\vx - \vx^\prime| } { \ell } \right) \frac { \Gamma ( p + 1 ) } { \Gamma ( 2 p + 1 ) } \sum _ { i = 0 } ^ { p } \frac { ( p + i ) ! } { i ! ( p - i ) ! } \left( \frac { \sqrt { 8 \nu } |\vx - \vx^\prime| } { \ell } \right) ^ { p - i }.
\end{align}
For this reason, we set $\nu=3/2$. The final covariance function, including the
$\sigma^2$ parameter to control the range of $f()$ therefore becomes
\begin{eqnarray}
k _ { \rvtheta}((x_{\text{E}}, x_{\text{N}}), (y_{\text{E}}, y_{\text{N}})) & = & \sigma^2 k _ { \ell, \nu = 3 / 2 } (x_{\text{E}}, y_{\text{E}}) \times k _ { \ell, \nu = 3 / 2 } (x_{\text{N}}, y_{\text{N}}),
\end{eqnarray}
where $\rvtheta = [\sigma^2, \ell]^\top$.

\subsection{Inference}
To infer posterior distribution of the regression coefficients, $\rvbeta$,
latent field $\rvf$, and its hyperparameters $\rvtheta$, we use a Hamiltonian
Monte Carlo sampler. The scale parameters $\sigma^2_1,\dots, \sigma^2_J$ are
analytically integrated out (see \eqref{eq:InferenceBetaPriorIntegrateOutSigma}
in the appendix). Due to positivity constraint of the hyperparameters, we sample
from $\rvphi = \log \rvtheta$ (applied component-wise). The density function of
the joint posterior distribution we are sampling from is proportional to the
product of likelihood and the priors, i.e.
\begin{equation}
p(\rvf, \rvbeta, \rvphi | \rvy) \propto p(\rvy | \rvf, \rvbeta) p(\rvf | \exp(\rvphi)) p(\rvbeta) p_{\rvtheta}(\exp(\rvphi)) \prod_i \left\vert \frac{d}{d\ervphi_i} \exp(\ervphi_i) \right\vert.
\end{equation}

To effectively use HMC sampler, log-likelihood of the posterior and its gradient
need to be tractable. Thanks to the grid structure of our study region, we
utilise Kronecker product structure that is present in the covariate matrix in
$p(\rvf|\rvtheta)$ if the covariance function $k_{\rvtheta}(\cdot,\cdot)$ is
assumed to be a product of covariance functions, one per each dimension (For
more details, see \citet{saatci_scalable_2012}). After expansion, the
unnormalised log-density becomes
\begin{eqnarray}
\log p(\rvf, \rvbeta, \rvphi | \rvy)
& = & \log p(\rvy | \rvf, \rvbeta) + \log p(\rvbeta) + \log p(\rvf | \exp(\rvphi)) + \log p_{\rvtheta}(\exp(\rvphi)) + \sum_i \ervphi_i +  \text{const}_1  \nonumber \\
& = & \left( \rvy^\top\mX\rvbeta + \rvy^\top \rvf - \exp(\mX\rvbeta + \rvf) \right) + \log p(\rvbeta) \nonumber \\
& & \quad + \left( - \frac{1}{2}\log \left| \mK_{\rvtheta} \right| - \frac{1}{2} \rvf^{\top} \mK_{\rvtheta}^{-1} \rvf \right) + \log p_{\rvtheta}(\exp(\rvphi)) + \sum_i \ervphi_i +  \text{const}_1,
\end{eqnarray}

The gradients of the log posterior density w.r.t. quantities of interest are
\begin{eqnarray}
\nabla_{\rvf} \log p(\rvf, \rvbeta, \rvphi | \rvy)
& = & \left( \rvy - \exp(\mX\rvbeta + \rvf)\right) + \left( - \mK_{\rvtheta}^{-1}\rvf \right)
\label{eq:method-mcmc-log-posterior-derivatives-field}\\
\nabla_{\rvbeta} \log p(\rvf, \rvbeta, \rvphi | \rvy)
& = & \left(\mX^\top \rvy - \mX^\top \exp(\mX\rvbeta + \rvf)\right) + \nabla_{\rvbeta} \log p(\rvbeta)
\label{eq:method-mcmc-log-posterior-derivatives-beta}\\
\nabla_{\ervphi_i} \log p(\rvf, \rvbeta, \rvphi | \rvy)
& = & \frac{1}{2}\rvf^\top \mK_{\rvtheta}^{-1} \frac{\partial \mK_{\rvtheta}}{\partial \ervtheta_i} \mK_{\rvtheta}^{-1} \rvf - \frac{1}{2} \mtrace\left(\mK_{\rvtheta}^{-1} \frac{\partial \mK_{\rvtheta}}{\partial\ervtheta_i}\right) \nonumber\\
& & \quad+ \nabla_{\ervphi_i} \log p_{\rvtheta}(\exp(\rvphi)) + 1.
\label{eq:method-mcmc-log-posterior-derivatives-hyperparams}
\end{eqnarray}

The expansion of un-normalised log-density of $\rvbeta$ and the gradients are
derived in \eqref{eq:InferenceBetaPriorLogDensity} and
\eqref{eq:InferenceBetaPriorNablaLogDensity} below.

All operations involving $\mK_{\rvtheta}$ can be sped up using Kronecker product
factorisation. Given $n^2$ is the number of elements in the full matrix
$\mK_{\rvtheta}$, operations in
\eqref{eq:method-mcmc-log-posterior-derivatives-field} and
\eqref{eq:method-mcmc-log-posterior-derivatives-hyperparams} can be computed in
$\BigOh{n^{\frac{3}{2}}}$ time by utilising the Kronecker structure in matrix
inversion and matrix-vector multiplication. For full details, see
\citet{saatci_scalable_2012}.

\section{Model derivations}
\label{sec:appendix_model_derivations}

\subsection{Beta prior}
Given a vector of $J$ independent random variables $\rvbeta$, of which each
component is distributed as follows
\begin{align}
\ervbeta_j & \sim  \mathcal{N}(0, \sigma_j^2), \nonumber\\
\sigma_j^2 & \sim  \text{InvGamma}(a, b). \nonumber
\end{align}
Let $\Psi = \left(\sigma_1^2, \dots, \sigma_J^2 \right)^\top $, then the prior
for the coefficients is given by integrating out the nuisance parameter $\Psi$
\begin{align}
\begin{split}
p(\rvbeta) & = \prod_j p(\ervbeta_j) \\
& = \prod_j \int p(\ervbeta_j | \Psi_j) p(\Psi_j)d\Psi_j  \\
& = \prod_j \int \frac{1}{\sqrt{2\pi}} \Psi_j^{-1/2}
\exp\left(-\frac{1}{2\Psi_j} \ervbeta_j^2 \right)
\frac{b^a}{\Gamma(a)} \Psi_j^{-a-1} \exp\left(-\frac{b}{\Psi_j}\right)   d\Psi_j \\
& = \prod_j \frac{b^a}{\sqrt{2\pi} \Gamma(a)} \int  \Psi_j^{-a-\frac{1}{2}-1}\exp\left(-\frac{\frac{1}{2} \ervbeta_j^2 + b}{\Psi_j}\right)    d\Psi_j \\
& = \prod_j \frac{b^a}{\sqrt{2\pi} \Gamma(a)} \frac{\Gamma\left(\frac{1}{2}
  + a\right)}{\left(\frac{1}{2}\ervbeta_j^2 + b\right)^{\frac{1}{2}+a}}
\end{split}
\label{eq:InferenceBetaPriorIntegrateOutSigma}
\end{align}

For the purposes of HMC, we derive both log-density and the gradient of
log-density w.r.t. the each individual components. Log-density is given as
\begin{align}
\begin{split}
\log p(\rvbeta) = \sum_i -\left(\frac{1}{2} + a\right)\log\left( \frac{1}{2}
\ervbeta_i^2 + b \right),
\end{split}
\label{eq:InferenceBetaPriorLogDensity}
\end{align}
from which the gradient is equal to
\begin{align}
\frac{\partial \log p(\rvbeta)}{\partial \ervbeta_i} & = \frac{(-\frac{1}{2} - a) \ervbeta_i}{\frac{1}{2} \ervbeta_i^2 + b}.
\label{eq:InferenceBetaPriorNablaLogDensity}
\end{align}

\subsection{Conditional densities for SAM-GLM inference}
The derivations below use the properties of the density function of the
Dirichlet distribution and the following property of the Gamma function,
$\Gamma(a + 1) = a\Gamma(a)$.

\subsubsection{Regression coefficients update}
\begin{align}
\begin{split}
p(\rvbeta | \alpha, \mX, \rvy, \rvz)
& \propto p(\rvy | \rvbeta, \mX, \rvz) p(\rvbeta) \\
& \propto \left\{ \prod_{k=1}^{K} \prod_{j=1}^{J} p(\ervbeta_{k,j})\right\}
\left\{ \prod_{n=1}^{N} p(\ervy_n | \rvbeta, \mX, \ervz_n)\right\} \\
& \propto \left\{ \prod_{k=1}^{K} \prod_{j=1}^{J} p(\ervbeta_{k,j})\right\}
\left\{ \prod_{n=1}^{N} \prod_{k=1}^{K} p(\ervy_n | \rvbeta_k, \mX)^{I(\ervz_n = k)}\right\} \\
& \propto \left\{ \prod_{k=1}^{K} \prod_{j=1}^{J} p(\ervbeta_{k,j})\right\}
\left\{ \prod_{n=1}^{N} \prod_{k=1}^{K}
\left(\frac{\exp(\mX_n^\top\rvbeta_k)^{\ervy_n}
  \euler^{-\exp(\mX_n^\top\rvbeta_k)}}{\ervy_n !}\right)^{I(\ervz_n =
  k)}\right\},
\end{split}
\label{eq:MixturesInferenceBetaSamplingFullDerivation}
\end{align}
where $p(\rvbeta)$ is expanded according to
\eqref{eq:InferenceBetaPriorIntegrateOutSigma}. For the purposes of Hamiltonian
Monte Carlo, the gradient of the posterior distribution is analytically
available.

\subsubsection{GPs updates}

The unnormalised joint posterior density of the $K$ GPs and their
hyperparameters is given as
\begin{align}
p(\mF, \rvtheta | \rvy, \rvz) & \propto p(\rvz | \mF) p(\mF|\rvtheta) p(\rvtheta) \nonumber \\
& \propto \prod_{n=1}^{N}p(z_n|\mF) \prod_{k=1}^{K}p(f_k|\rvtheta_k)p(\rvtheta_k) \nonumber \\
& \propto \prod_{n=1}^{N}\prod_{k=1}^{K}\left(\frac{\exp(\ervf_{k,b[n]})}{\sum_{l=1}^{K}\exp(\ervf_{l,b[n]})}\right)^{I(z_n=k)} \prod_{k=1}^{K}p(f_k|\rvtheta_k)p(\rvtheta_k),  \nonumber
\end{align}
where $p(f_k|\rvtheta_k)$ is the density function of the zero-mean multivariate
Gaussian distribution with covariance matrix parameterised by $\rvtheta$, and
$p(\rvtheta_k)$ is a suitable prior for the hyperparamers. The gradient of the
joint posterior with respect to $\mF$ and $\rvtheta$ are analytically available.

\subsubsection{Mixture allocation update for spatially-dependent blocks}
\label{sec:appenidx_mixture_allocation_update_for_dependent_blocks}
\begin{align}
p(\ervz_n = k | \rvz^{\bar{n}}, \mX_n, \rvbeta, \rvy, \mF)
& = p(\ervy_n|\ervz_n=k, \mX_n, \rvbeta_k)p(z_n | \mF) \nonumber \\
& \propto p(\ervy_n|\ervz_n=k, \mX_n, \rvbeta_k)\frac{\exp(\ervf_{k,b[n]})}{\sum_{l=1}^{K}\exp(\ervf_{l,b[n]})} \nonumber\\
& = \prod_{k=1}^{K} \left(\frac{\exp(\mX_n^\top\rvbeta_k)^{\ervy_n} \euler^{-\exp(\mX_n^\top\rvbeta_k)}}{\ervy_n !}\right)^{I(\ervz_n = k)}\frac{\exp(\ervf_{k,b[n]})}{\sum_{l=1}^{K}\exp(\ervf_{l,b[n]})} \nonumber
\end{align}

\subsubsection{Mixture allocation update for independent blocks}
\label{appendix:sssec_independent_weights_mixture_allocation_update}
\begin{align}
p(\ervz_n = k | \rvz^{\bar{n}},\alpha, \mX_n, \rvbeta, \rvy)
& \propto p(\ervy_n | \ervz_n=k, \mX_n, \rvbeta) \int p(\ervz_n | \rvpi_{b[n]}) p(\rvpi_{b[n]} | \alpha, \rvz^{\bar{n}}) d \rvpi_{b[n]} \nonumber\\
& \propto p(\ervy_n | \ervz_n=k, \mX_n, \rvbeta) \int \prod_k \ervpi_{b[n], k}^{I(\ervz_n = k)}     \frac{\Gamma(\sum_{j=1}^{K} B_{b[n],j})}{\prod_{j=1}^{K} \Gamma(B_{b[n],j})} \prod_{j=1}^{K}\ervpi_{b[n],j}^{B_{b[n],j}-1}      d \rvpi_{b[n]} \nonumber \\
& \propto p(\ervy_n | \ervz_n=k, \mX_n, \rvbeta) \frac{\Gamma(\sum_{j=1}^{K} B_{b[n],j})}{\prod_{j=1}^{K} \Gamma(B_{b[n],j})} \frac{\prod_{j=1}^{K} \Gamma(B_{b[n],j} + I(j=k))}{\Gamma(\sum_{j=1}^{K} B_{b[n],j} + I(j=k))} \nonumber \\
& \propto p(\ervy_n | \ervz_n=k, \mX_n, \rvbeta)   \frac{B_{b[n], k}}{ \sum_{j=1}^{K} B_{b[n],j}}       \nonumber \\
& \propto \prod_{k=1}^{K} \left(\frac{\exp(\mX_n^\top\rvbeta_k)^{\ervy_n} \euler^{-\exp(\mX_n^\top\rvbeta_k)}}{\ervy_n !}\right)^{I(\ervz_n = k)} \frac{c_{b[n]k}^{\bar{n}} + \alpha}{ K \alpha + \sum_{j=1}^{K} c_{b[n]j}^{\bar{n}} }, \label{eq:independent_weights_mixture_allocation_update_full_derivation}
\end{align}
where $B_{b,k} = c_{b,k}^{\bar{n}} + \alpha$, and $c_{b,k}^{\bar{n}}$ is the
number of cells in block $b$ other than cell $n$ that are assigned to component
$k$.

\section{Dependence of blocks -- extra plots}
\label{sec:appendix-block-dependence}

This section includes two plots related to the discussion of dependence of
blocks in the paper. We compare the independent blocks version of our model with
the variant that addresses the dependence via Gaussian random fields. The plots
below show that considering dependence between the blocks can improve model
predictions in some cases but it requires sampling from a high-dimensional
distribution ($K \times B$), resulting in slow mixing.

\begin{figure}
  \centering%
  \includegraphics[width=\textwidth]{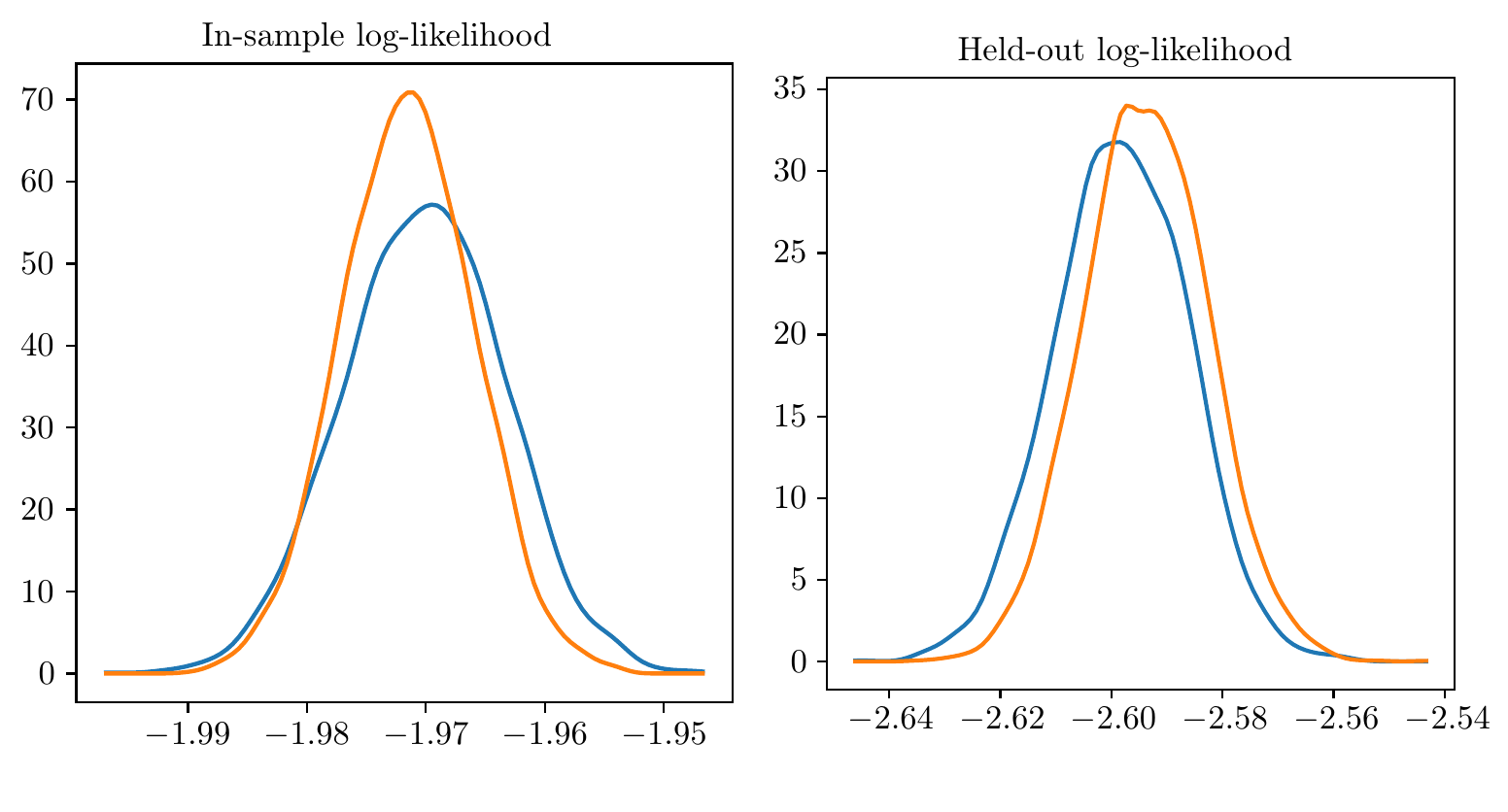}
  \caption{\label{fig:dep-vs-indep-smoothed-histogram}Smoothed histograms of log
    likelihood computed on in-sample counts (left), and out-of-sample counts
    (right) using the proposed model with dependent blocks (\LineOrange), and
    independent blocks (\LineBlue) when $K=3$. Blocking: MSOA, training data:
    2015, test data: 2016, model specification 4.}
\end{figure}

\Figref{fig:dep-vs-indep-smoothed-histogram} compares smoothed histograms for
samples of in-sample log-likelihood $p(\rvy | \rvphi)$ for both variants of the
model when $K=3$, with their out-of-sample counterpart using samples from
$p(\tilde{\rvy}|\rvphi)$. While independent-blocks model performs better
in-sample, the dependent-blocks model generalises better to out-of-sample
data. However, for $K=2$ and $K=4$, the model with independent blocks has lower
RMSE on out-of-sample data as reported in the paper.

\begin{figure}
  \centering%
  \includegraphics[width=\textwidth]{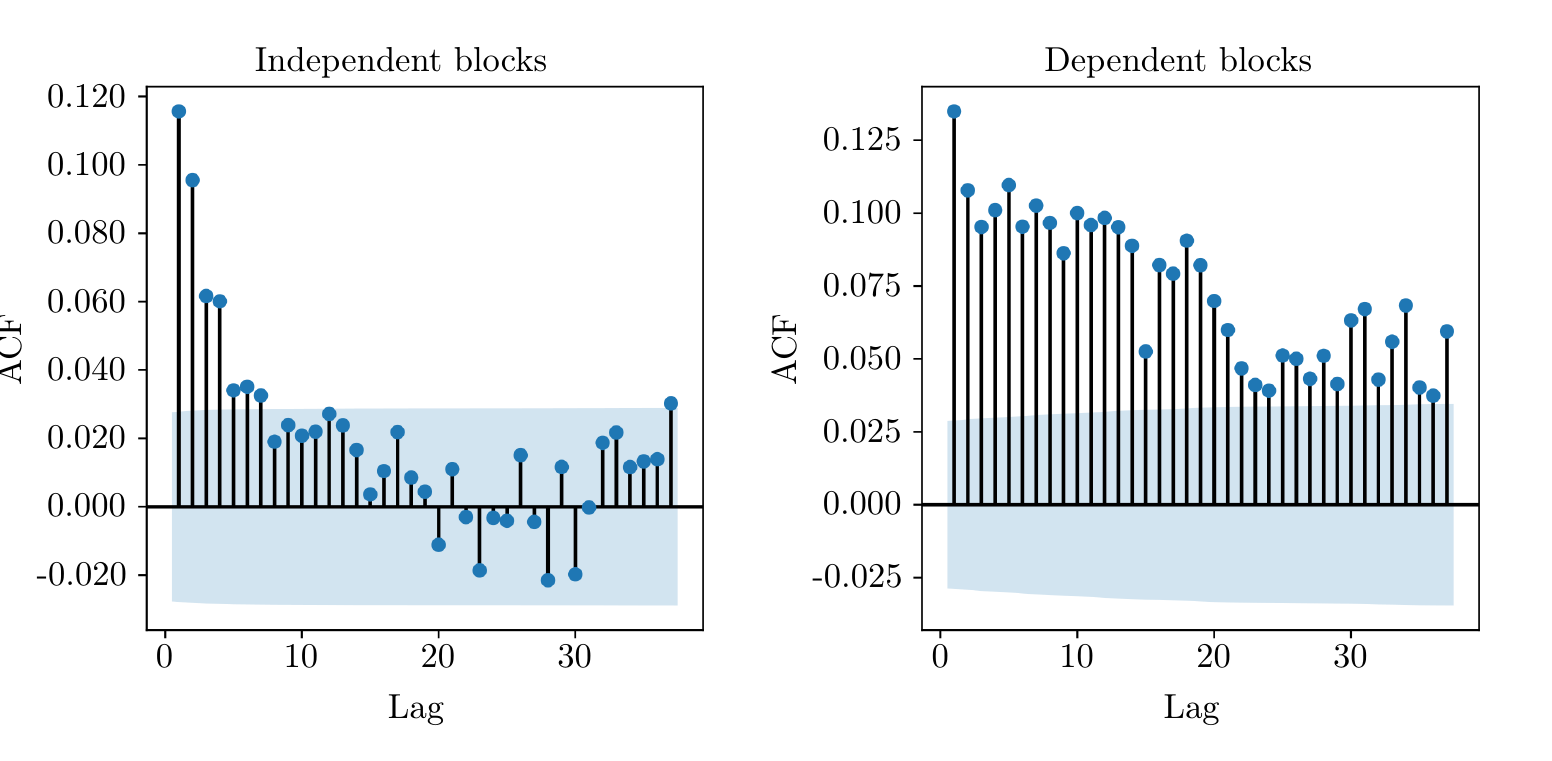}
  \caption{\label{fig:dep-vs-indep-acf-plots-comparison}Autocorrelation plots
    for the samples of in-sample log-likelihood when $K=3$. Blocking: MSOA,
    training data: 2015, model specification 4.}
\end{figure}

\Figref{fig:dep-vs-indep-acf-plots-comparison} shows the autocorrelation plot
for the in-sample log-likelihood obtained from \num{50000} samples that were
thinned to \num{5000} for both variants to assess mixing performance. It is
clear that successive samples obtained from the complex dependent-blocks model
are more correlated to each other than for the case of independent blocks
indicating slower mixing. Further, the inferences made using a Markov chain with
high autocorrelation may lead to biased results.

\end{document}